\documentclass{aa}
\usepackage[comma,authoryear]{natbib}
\usepackage{graphics}
\usepackage[latin1]{inputenc} 

\begin{document}

 \title{A new method of correcting radial velocity time series for inhomogeneous convection}

   \titlerunning{}

   \author{N. Meunier \inst{1,2}, A.-M. Lagrange \inst{1,2}, S. Borgniet \inst{1,2}  
          }
   \authorrunning{Meunier et al.}

   \institute{
Univ. Grenoble Alpes, IPAG, F-38000 Grenoble, France\\
CNRS, IPAG, F-38000 Grenoble, France\\
  \email{nadege.meunier@univ-grenoble-alpes.fr}
             }

\offprints{N. Meunier}

   \date{Received 22 December 2016; Accepted 18 July 2017}

\abstract{Magnetic activity strongly impacts stellar radial velocities ({\it RVs}) and therefore  the search for small planets. We  showed previously that in the solar case it induces $RV$ variations with an amplitude over the cycle on the order of 8~m/s, with signals on both short and long  timescales. The major component is the inhibition of the convective blueshift due to plages. }
{In this paper we explore a new approach used to correct for this major component of stellar radial velocities in the case of solar-type stars. }
{The convective blueshift depends on line depths; we use this property to develop a method that will  characterize  the amplitude of this effect and to correct for this $RV$ component. We build realistic $RV$ time series corresponding to $RVs$ computed using different sets of lines, including lines in different depth ranges. We characterize  the performance of the method used to reconstruct the signal without the convective component and the detection limits derived from the residuals.}
{We identified a set of lines which, combined with a global set of lines, allows us to reconstruct the convective component with a good precision and to correct for it. For the full temporal sampling, the power in the range 100-500~d  significantly decreased, by a factor of 100  for a $RV$ noise below 30~cm/s. We also studied the impact of  noise contributions other than the photon noise, which  lead to uncertainties on the $RV$ computation, as well as the impact of the temporal sampling. We found that these other sources of noise do not greatly alter  the quality of the correction, although they need a better noise level to reach a similar performance level. }
{A very good correction of the convective component can be achieved providing very good $RV$ noise levels combined with a very good instrumental stability and realistic granulation noise. Under the conditions considered in this paper, detection limits at 480~d lower than 1 M$_{\rm Earth}$ could be achieved for  $RV$ noise below 15~cm/s. }


\maketitle

\section{Introduction}

Stellar variability at various timescales strongly affects the ability to detect exoplanets. 
The magnetic activity contribution to radial velocities ({\it RVs}) is due to the following components \cite[][]{meunier10a}:  the photometric contribution of spots, plages, and network (hereafter $RV_{\rm sppl}$), which depends on their intensity contrast and size, and the  attenuation of the convective blueshift in plages (hereafter $RV_{\rm conv}$), which depends on the attenuation of the convective blueshift and plage size. 
In the case of the Sun, the latter is expected to dominate the signal, as shown in Fig.~\ref{rv}. 
Attempts to correct for the $RV_{\rm conv}$ signal have been made using different techniques: correlation with chromospheric emission \cite[][]{meunier13}, which provides correction on both long (cycle) timescales and short (rotational) timescales, or correlation with a smoothed chromospheric emission \cite[][]{dumusque12} to remove some contribution on long timescales; harmonic fittings or fits using a limited number of structures to remove some stellar signals at the rotational period  \cite[e.g.,][]{boisse11,dumusque12,dumusque14}; use of photometric times series to estimate the {\it RV} signal \cite[][]{aigrain12}.

On the other hand, it has been shown that the amount of convective blueshift, when the spectral line positions are computed using the bottom of lines, i.e., the lower part of the line around the line center, and eliminating the contribution of the wings depends on the depths of the spectral lines used to compute the {\it RV} \cite[][]{dravins81}, controlling directly the $RV_{\rm conv}$ amplitude, while $RV_{\rm sppl}$ does not depend on these line depths. 
We propose to use that property to retrieve the different components from several {\it RV} time series computed with different sets of lines and attempt to correct the observed {\it RV} for the convective component. The differential velocity shifts of spectral lines, which correspond to the velocity shifts computed for various spectral lines versus the line depth \cite[see][for a discussion about the difference between the relative and absolute shift]{meunier17}, have been studied for the Sun and small samples of stars \cite[][]{dravins81,gray82,dravins87b,dravins99,hamilton99,landstreet07,allendeprieto02,gray09}.  \cite{meunier17} have studied this effect for a much larger sample of stars (167 main sequence G and K stars using HARPS spectra) and showed for the first time the impact of magnetic activity on it. \cite{reiners16} have also recently reevaluated precisely this signature for the Sun.  

Our objective is to test the performance of a correction method based on the computation of two different {\it RV} time series from the same observed spectra, but using different spectral lines for different noise levels on {\it RV}. We focus on stars with a convection amplitude similar to that of the Sun. 
The outline of the paper is the following. 
In Sect.~2, we present the method. 
The results are described in Sect.~3: we characterize the reconstructed time series and and evaluate the performance of the correction. We study the impact of the temporal sampling and of our assumptions on our results in Sect.~4, and test our method on  current HARPS data. We conclude in Sect.~5.

\section{Method}

\subsection{Philosophy of our approach}

\subsubsection{General principles}

\begin{figure}
\includegraphics{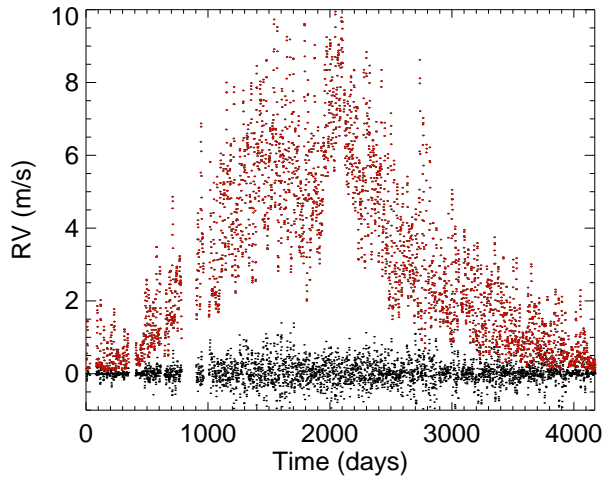}
\caption{
{\it RV} due to spots and plages (black) and convection attenuation in plages (red) in the solar case, from Meunier et al. (2010). }
\label{rv}
\end{figure}

Measured {\it RVs} are the sum of several contributions: the {\it RV} due to the attenuation of the convective blueshift, the {\it RV} due to the photometric contribution of spot and plages, and the {\it RV} due to other sources impacting short timescales such as granulation and photon noise.
Radial velocity time series computed using different sets of spectral lines corresponding to different depths should exhibit a different amplitude because the convective blueshift induced contribution depends on the line depth.
The measured {\it RV} is therefore the sum of two types of {\it RV}, one (including $RV_{\rm sppl}$) is independent of the lines used to compute {\it RV}, while the other  depends on the choice of spectral lines. 

In the following, we focus on three components: the photometric contribution of spots and plages, the convective component due to inhomogeneous magnetic activity, and photon noise (which is modeled by a Gaussian noise applied to the {\it RV}  time series). We call $RV_{\rm conv}$ the convective contribution which would be obtained when using a large set of spectral lines S$_0$. The same convective component but measured with another set of spectral lines is $\alpha RV_{\rm conv}$, where $\alpha=\Delta V /\Delta V_0$ is the ratio between the convective blueshift corresponding to that set of lines and the convective blueshift corresponding to S$_0$.
Because a given set of lines uses only a subset of the lines present in the spectra, given a certain signal-to-noise ratio (S/N) on the spectra the uncertainties on the computed {\it RV} differ from one set of lines to the other. We study these properties for the different sets of lines and test different methods for retrieving the two components (spot+plage and convection) from different time series.

\subsubsection{Outline of the method}

The problem to solve can then be described as follows. A time series  $RV_0(t)$ is computed from  a large set of lines
S$_0$, while another time series $RV_1(t)$ is computed from a set of lines S$_1$ including only lines with flux within a restricted  range
for which the convective blueshift is different from that due to S$_0$,
\begin{eqnarray}
RV_0(t) & = & RV_{\rm sppl}(t) + RV_{\rm conv}(t) \\
RV_1(t) & = & RV_{\rm sppl}(t) + \alpha_1 RV_{\rm conv}(t) \,,
\end{eqnarray}

where $\alpha_1=\Delta V_1 /\Delta V_0$ is the ratio between the blueshifts corresponding to the two sets of lines. We recall that  $RV_{\rm sppl}$ is the photometric contribution of spots and plages to {\it RV}, and $RV_{\rm conv}$ is the contribution to {\it RV} due to the attenuation of the convective blueshift in plages.
We neglect the chromatic effect on $RV_{\rm sppl}$ here (because we consider a relatively small range in wavelength).
The question is then is it possible to retrieve the $RV_{\rm sppl}$ and $RV_{\rm conv}$ time series from the $RV_0$ and $RV_1$ time series,  and if so with what precision?
From  a mathematical point of view, if $\alpha_1$ is known, it is straightforward to solve this system of equations for each time step, while if $\alpha_1$ is not known some assumptions must be made in order to solve them.

\begin{table}
\begin{center}
\caption{Reference series properties}
\label{tabref}
\begin{tabular}{cccccc}
\hline \hline
Series  & rms {\it RV}    &  Long-term & average & Minimum & Maximum\\
        &           &   amplitude &        &         &         \\   
$RV_{\rm sppl}^t$ &  0.33 &  0 &  0.02 & -2.42 & 2.19 \\
$RV_{\rm conv}^t$ &  2.38 & 8.2 &  3.17 & 0.09 & 10.08 \\
\hline
\end{tabular}
\tablefoot{Values are in m/s. The reference series are those derived by \cite{meunier10a} from observed structures on the Sun during a solar cycle.
 }
\end{center}
\end{table}

We use the $RV_{\rm sppl}$ and $RV_{\rm conv}$ (which we wish to correct for in this paper) obtained by \cite{meunier10a} as reference series. They are considered  ``true'' series, and  we will attempt to retrieve them; hereafter they are denoted $RV_{\rm sppl}^t$ and $RV_{\rm conv}^t$. Table~\ref{tabref}
summarizes important properties of these time series. Then, we implement the following procedure:

\begin{itemize}
\item{
{\it Step 0: Characterizing and choosing the best sets of lines.} We define the sets of lines and their properties: this defines      $\Delta V$ hence $\alpha$ for each set of lines, and the noise on {\it RV} (Sect.~2.2);}
\item{{\it Step 1: Building synthetic {\it RV} time series corresponding to the different sets of lines.} 
We use $RV_{\rm sppl}^t$ and $RV_{\rm sppl}^t$ to build the synthetic time series $RV_0$ and $RV_1$ (corresponding to two sets of lines S$_0$ and S$_1$) according to Eqs. 4 and 5, using $\alpha$ and some specific noise for each measurement accordingly (Sect.~2.3);}
\item{{\it Step 2:  Choosing the value of $\alpha$ and retrieving reconstructed series $RV_{\rm sppl}^r$ and $RV_{\rm conv}^r$}. The value of $\alpha$ is either known (precisely or with some uncertainty) or we must estimate it from the $RV_0$ and $RV_1$ series. The system is then solved under various assumptions (with no a priori knowledge on the value of $\alpha$), leading to reconstructed $RV_{\rm sppl}^r(t)$, $RV_{\rm conv}^r(t)$ and $\alpha$ (see Sect.~2.4); }
\item{{\it Step 3: Testing the quality of the reconstruction.} These reconstructed values ($RV_{\rm sppl}^r(t)$, $RV_{\rm conv}^r(t)$, and $\alpha$) are compared to the input values from Step 1 (Sect.~2.5); }
\item{{\it Step 4: Applying a correction to $RV_0$. } The simulated series can be corrected for the convective component by subtracting $RV_{\rm conv}^r$ from $RV_0$ (Sect.~2.6);}
\item{{\it Step 5: Testing the quality of the {\it RV} correction.} The residuals after correction are analyzed and characterized (Sect.~2.6).}
\end{itemize}

\subsection{Step 0: Line set determination and properties}

\subsubsection{Sets of lines}

\begin{figure}
\includegraphics{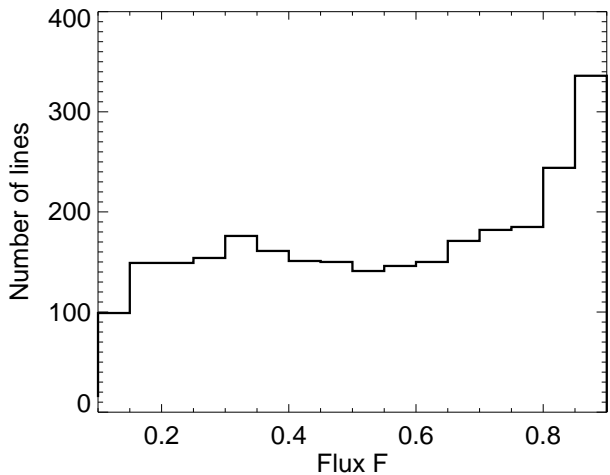}
\caption{
Distribution of the line fluxes in the reference line set S$_0$.
}
 \label{histo}
\end{figure}

To determine the line depth, we use the solar optical spectra from \cite{kurucz84} and re-reduced in 2005 by Kurucz \footnote{http://kurucz.harvard.edu/sun/fluxatlas2005/}.
We identify all lines with a flux $f$ (at the bottom of the lines) between 0.05 and 0.9 for wavelengths between 4000 and 6600~\AA, producing a line set used as a reference. This leads to a set of 3858 lines, constituting the reference set of lines S$_0$. Figure~\ref{histo} shows the distribution of the fluxes for these lines.  From S$_0$ we can also select lines with fluxes between $F_1$ and $F_2$, forming new sets of lines: a set of lines is defined by the selection of lines with flux between a minimum flux $F_1$ and a maximum flux $F_2$.

\subsubsection{Set of line properties: $\Delta V$ and $P$}

\begin{figure}
\includegraphics{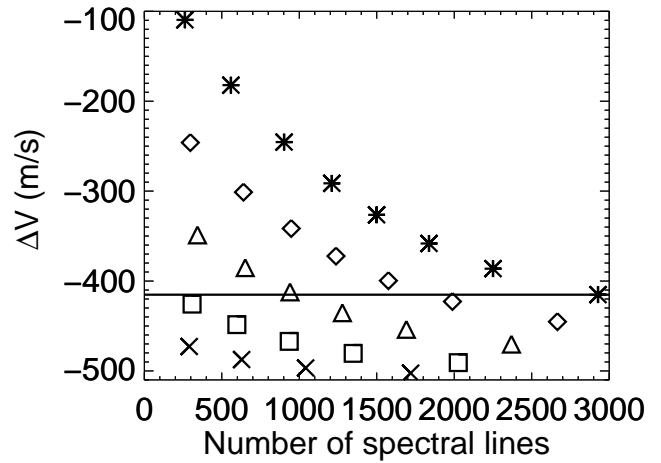}
\caption{
$\Delta V$ versus the number of lines for various sets of lines for a minimum flux of 0.05 (stars), 0.2 (diamonds), 0.3 (triangles), 0.4 (squares), and 0.5 (crosses); the maximum varies between a value above the minimum up to 0.9. The horizontal line corresponds to S$_0$. }
\label{dv1}
\end{figure}

\begin{figure}
\includegraphics{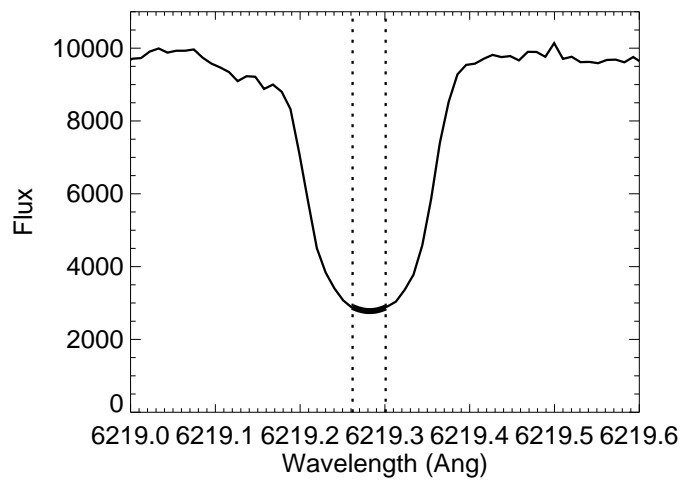}
\caption{
Example of a spectral  line (thin solid line) and the second-degree polynomial fit around line center (thick solid line) delimited by the two vertical dotted lines.
}
\label{fitline}
\end{figure}

For a given set of spectral lines, we estimate a realistic $\Delta V$ as follows. 
We compute the convective blueshift associated with each spectral line using the relationship obtained by \cite{reiners16} for the Sun between the shift of an individual line $\delta V_i$ and the line depth $x_i=1-F_i$:
  
\begin{eqnarray}
\delta V_i=-504.891-43.7963 x_i-145.560 x_i^2+884.308 x_i^3 \,.
\end{eqnarray}
The average  of $\delta V_i$ over the required set of lines provides the corresponding $\Delta V$.  
Figure~\ref{dv1} illustrates the typical values taken by $\Delta V$ for thirty different sets of lines as a function of the number of lines identified in that set. 
This is discussed in Sect.~2.2.4. The different sets illustrated here correspond to $F_1$ with values of 0.05, 0.2, 0.3, 0.4, and 0.5, and $F_2$ with values between $F_1$+0.1 and 0.9. Each set therefore includes a different number of spectral lines (which is not chosen a priori).

\subsubsection{Uncertainties of computed {\it RV} for different sets of lines}

\begin{figure}
\includegraphics{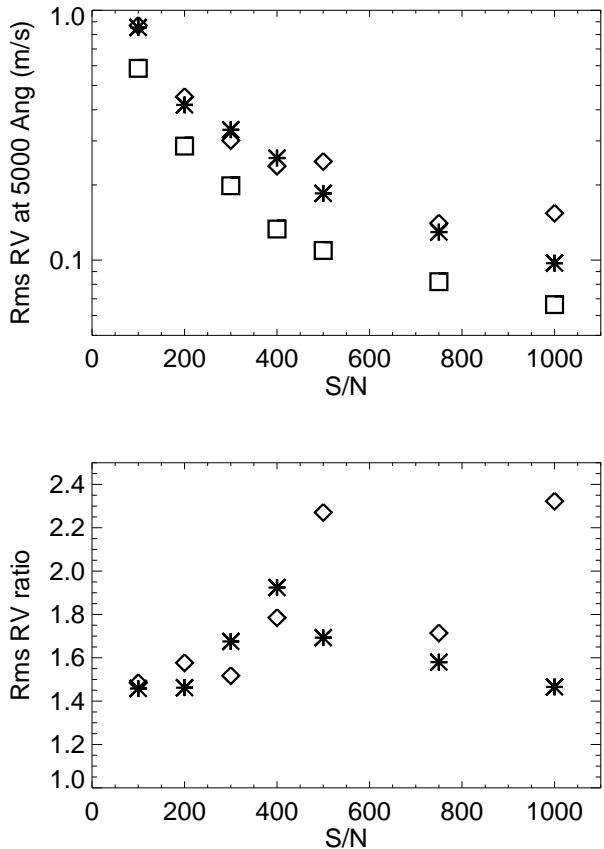}
\caption{
Rms {\it RV} versus S/N for set S$_0$ (0.05-0.9, squares), S$_1$ (0.05-0.5, stars), and S$_2$ (0.5-0.9, diamonds). 
}
\label{rms_sn}
\end{figure}

We use here the synthetic solar optical spectra used in the SAFIR software \cite{galland05} from \cite{kurucz93}, as in our previous simulations \cite[][]{desort07,lagrange10b,meunier10a,borgniet15}.
The SAFIR software computes {\it RV} from cross-correlations between spectra \cite[][]{chelli00}, and can be applied to  observed stellar spectra \cite[e.g.,][]{galland05} and also to simulated spectra such as in \cite{desort07}, \cite{lagrange10b}, or \cite{meunier10a}.  This spectrum, with a pixel size of 0.0063 \AA, has been convolved with the HARPS instrumental response \cite[in practice a convolution by a Gaussian whose full width at half maximum is the instrumental resolution;][]{mayor03} and the continuum is equal to 1.

For a given set of lines and  S/N on each pixel of the spectra, the computation of the shift between two spectra for many realizations of the  photon noise on the spectra  
provides a series of {\it RVs} whose root mean square (hereafter rms) gives the uncertainty on the resulting {\it RVs} due to the photon noise. This is performed as follows. 
For each set of lines, we add the corresponding photon noise to the synthetic spectra (for a given S/N $y$, the spectra is multiplied by $y^2$, a noise equal to the square root of the intensity at each pixel is then added: the indicated S/N therefore corresponds to the continuum, while the S/N is therefore larger at the bottom of the lines  where the flux is lower). One hundred realizations of the noise are performed. The average spectra is computed and is used as a reference. The bottom of the line positions are computed for this reference spectra and for each of the 100 realizations for each line in the set using a second-degree polynomial fit over $\pm0.02$~$\AA$, the difference between the two providing a {\it RV} for that realization. Such a  fit is illustrated in Fig.~\ref{fitline}. The choice of 0.02~$\AA$ is a compromise between selecting enough points to be able to perform the polynomial fit and the need to consider only the center of the lines. The rms {\it RV} over the 100 realizations gives the uncertainty corresponding to that set of lines and the S/N.  
The square symbols in Fig.~\ref{rms_sn} shows the uncertainties versus S/N for the set of lines S$_0$: it reaches the 10~cm/s level for S/N around 2000.

For simplicity we consider only the {\it RV} uncertainty related to the {\it RV} computation, which is directly related to the  S/N on the spectra and therefore to the photon noise. However, it does not include the {\it RV} uncertainty related to the instrumental stability for example, which would take the same value for all sets of lines (see Sect.~4.3 for a discussion on this issue).

\begin{figure}
\includegraphics{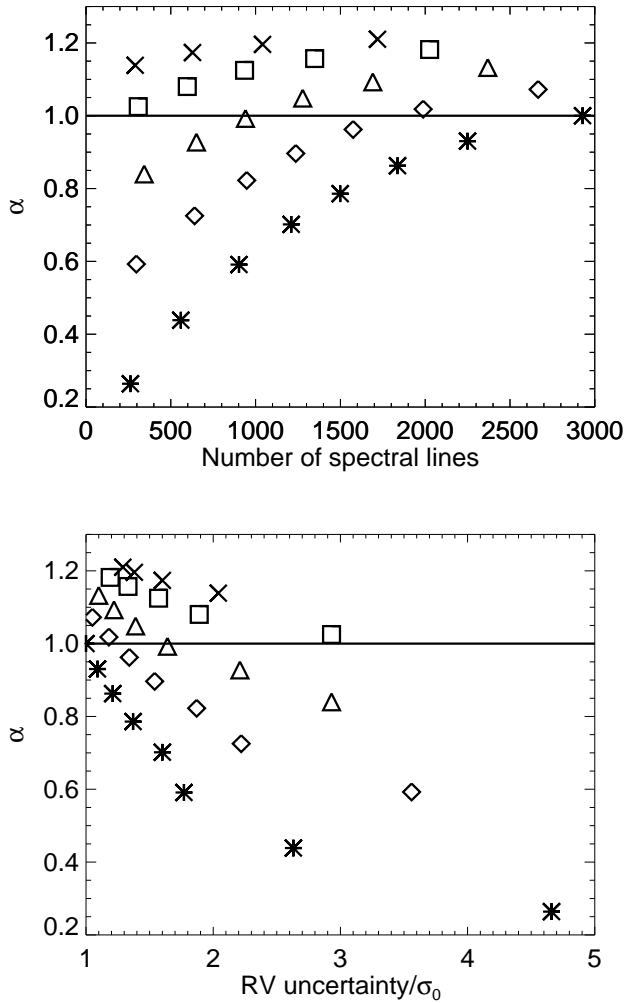}
\caption{
{\it Upper panel:} $\alpha$ versus the number of lines for various sets of lines (same symbol code as Fig.~3). 
{\it Lower panel:} Same, but for $\alpha$ versus the uncertainty ratio (rms {\it RV} for the set of lines divided by the rms {\it RV} for S$_0$). 
}
\label{dv2}
\end{figure}

\subsubsection{Line set choice}

Fig.~\ref{dv1} and Fig.~\ref{dv2} illustrate the typical values taken by $\Delta V$ and $\alpha$
for different sets of lines, as a function of the number of lines 
identified in that set. We note that at this stage $\alpha$ depends only on the $\Delta V$ estimated in the previous section, not on the $RV^t(t)$ series. The value of 
$\alpha$ is also shown as a function of the 
ratio $R$ defined as the ratio between the rms {\it RV} for the considered set of lines and the rms {\it RV} for S$_0$. This allows in the following  the uncertainties to be expressed as a function of the uncertainties derived for S$_0$, closely related to the usual uncertainties in the literature: for example, the usual {\it RV} computation techniques \cite[e.g.,][]{galland05} for HARPS  use all lines available with associated uncertainties corresponding to this set of lines. The amplitude depends on the S/N and on the spectral type, but the usual S/N for solar-type stars in the ESO archives is in the range 0.5-1~m/s. To obtain the best reconstruction in the next sections, we know that we  need to compute {\it RV} time series that are as different as possible with the best noise levels, and therefore to choose a set of lines with
\begin{itemize}
\item{$\alpha$ as far from 1 as possible;}
\item{$R$ (or rms {\it RV}) as low as possible.}
\end{itemize}

We note that the uncertainties on $\Delta V$ determined in Sect.~2.2.3 are in the range of a few m/s, i.e., they significantly differ from each other, which should lead to {\it RV} time series that will differ sufficiently from each other. In the following $\Delta V$ (or $\alpha$) is used as input, which we are attempting to retrieve with as little a priori knowledge as possible; therefore, the exact uncertainties are not important.

It should be noted that stars with identical spectral types but different levels in small-scale convection such as granulation (either on average or its temporal variability) impacting the convective blueshift, such as that derived by \cite{meunier17}, will give a similar $\alpha$ if the differential velocity shift of spectral lines is universal, as pointed out by \cite{gray09}, because the shape  of the differential velocity shift will be similar to that of Eq. 3. However, the value of $\alpha$ will  vary for a given set of lines from one spectral type to the other because the same lines correspond to a different flux range as eq. 3 is not linear. However, this will not be strongly affected by the level of convection itself because $\alpha$ is a relative variable. Our method is based the strong variability of the convective signal with time due to inhomogeneity from plages: it cannot be used if the star is quiet or when considering a large number of points over a very short time (for example one night).

\begin{figure}
\includegraphics{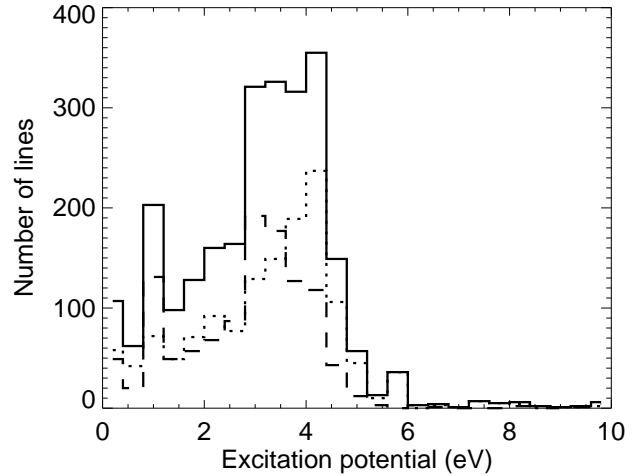}
\caption{
Distribution of the excitation potentials for 2532 lines of S$_0$ (solid line), 1146 lines of S$_1$ (dashed line), and 1386 lines of S$_2$ (dotted line). Line identifications were made using the spectrum of \cite{wallace07} and the solar spectrum available in the BASS2000 archive (http://bass2000.obspm.fr/),  and the excitation potentials were retrieved from the VALD archive \cite[][]{piskunov95,Rya99,kupka99,kupka00,Rya15}.
}
\label{potexc}
\end{figure}

We identify two sets of lines corresponding to different compromises between the two constraints, with fluxes in the range 0.06-0.5 (S$_1$) and 0.5-0.9 (S$_2$).
The rms {\it RV} versus $\sigma_0$ (hereafter the uncertainty on {\it RV} for S$_0$, $\sigma_0$ is on the order of 0.5-1 m/s for current observations with HARPS using cross-correlation techniques with reference masks or reference spectra) is shown in Fig.~\ref{rms_sn} for two sets of lines S$_1$ and S$_2$ and compared to S$_0$.  The ratio between the rms {\it RV} for S$_1$ (or S$_2$) with the rms {\it RV} for S$_0$ will be used in the following simulations to estimate the noise on each time series, given a {\it RV} noise level for S$_0$. The properties of the sets used in the following  are shown in Table~\ref{tabmask} (the ratio $R$ has been averaged over the eight S/N levels illustrated in Sect.~2.2.3 and Fig.~\ref{rms_sn}). In the next sections we  focus on the results obtained with the set of lines S$_1$. Its performance level is very similar (although marginally better) to that obtained with S$_2$.  In addition, for most of the considered S/N values, S$_2$ shows properties  between the two other sets, and the difference between sets here is probably due to the number of lines in each of them (S$_2$ includes more lines than S$_1$).
Figure~\ref{potexc} shows the distribution of the excitation potential for the different sets of lines for a large fraction of spectral lines. Although the dispersion is large \cite[][]{chiavassa11}, the average excitation potential is lower for S$_1$ (2.80) than for S$_2$ (3.22). 

\begin{table}
\begin{center}
\caption{Selected sets of line properties}
\label{tabmask}
\begin{tabular}{ccccccc}
\hline \hline
Set of &  $F_1-F_2$     &  $\Delta V$   & $\alpha $ &  $<F>$   &  Number  & S/N \\ 
lines     &               &  (m/s) &  &  & of lines &  \\ \hline
S$_0$    &  0.05-0.9   & -415.2  & -  & -  & 2899  &  - \\
S$_1$    &  0.05-0.5   & -291.4  & 0.70   & 0.309  & 1199  & 1.60  \\
S$_2$    &  0.5-0.9    & -502.5  & 1.21   & 0.739  & 1701   & 1.29  \\ \hline
\end{tabular}
\tablefoot{
$F_1-F_2$ is the line flux considered for the set of lines. 
The S/N  is the ratio $R$ between the rms {\it RV} for the considered set of lines and the rms {\it RV} for S$_0$. 
 }
\end{center}
\end{table}

\subsection{Step 1: building of the time series for S$_0$ and S$_1$}

We build {\it RV} time series as follows for S$_0$ and S$_1$ respectively: 

\begin{eqnarray}
RV_0(t) & = & RV_{\rm sppl}^t(t) + RV_{\rm conv}^t(t)+b_0(t) \\
RV_i(t) & = & RV_{\rm sppl}^t(t) + \alpha_i^t RV_{\rm conv}^t(t)+b_i(t) \,,
\end{eqnarray}
where $b_i$ is the noise due to the {\it RV} computation added to each {\it RV} time series and the exponent ``t'' indicates the reference values. 
For S$_0$, $b_0$ has a rms of $\sigma_0$, which varies between 0 and 1 m/s with  steps of 0.01 m/s.
For S$_1$, the rms of $b_i(t)$ is $R$ (defined in Sect.~2.2.4 and Tab.\ref{tabmask}) times $\sigma_0$, but the two time series $b_0(t)$ and $b_i(t)$ are not correlated.
Twenty realizations of the noise are performed for each noise level. 

\subsection{Step 2: Choice of $\alpha$ and reconstructed {\it RV} time series}

We consider two cases to estimate $\alpha$:  
\begin{itemize}
\item{{\it Case 1: $\alpha$ is known independently with some uncertainty}. 
We characterize the quality of the reconstructed $RV_{\rm sppl}^r$ 
and $RV_{\rm conv}^r$ for a given uncertainty on $\alpha$, the exponent ``r'' indicating reconstructed values;}
\item{{\it Case 2: $\alpha$ is not known at all}. This is the most general case. We therefore must make assumptions to solve the system in order to estimate $\alpha$ from our {\it RV} time series. 
}
\end{itemize}

\subsubsection{Case 1: $\alpha$ known with a given uncertainty}

For a given $\alpha$, the system described by eqs. 4 and 5 for sets of lines S$_0$ and S$_1$ can be solved to provide the reconstructed {\it RV} times series:

\begin{eqnarray}
RV_{\rm conv}^r(t) & = & (RV_1(t)-RV_0(t))/(\alpha-1) \\
RV_{\rm sppl}^r(t) & = & RV_0(t)-RV_{\rm conv}^r(t) \,;
\end{eqnarray}

Here we consider that $\alpha$ may be known with a certain uncertainty;
$\alpha$ could indeed be estimated independently from the {\it RV} series, for example by analyzing the spectra, as done by \cite{gray09,meunier17}, either for the star being studied or for the spectral type corresponding to it, or by magnetohydrodynamic numerical simulations of convection associated with the production of spectra  for various spectral types \cite[such as those produced  by e.g.,][]{ramirez09,chiavassa11,allendeprieto13,magic14} that would then be analyzed as observed spectra. Such techniques to derive $\alpha$, independent of the {\it RV} time series, have not yet been fully developed, but may in the future allow for a complementary computation of $\alpha$. It should be noted that if the differential velocity shift of spectral lines is universal, as claimed by \cite{gray09}, we expect the ratio $\alpha$ to vary little from one star to the next, because lines tend to be deeper for lower mass (main sequence) stars. 
We solve the equations for two values, $\alpha-\sigma_\alpha$ and $\alpha+\sigma_\alpha$, to provide a reconstruction of the {\it RV} time series in two extreme conditions;
$\alpha$ is the true value and $\sigma_\alpha$ is the typical uncertainty on $\alpha$.

\subsubsection{Case 2: $\alpha$ unknown}

For a given star, the value of $\alpha$ is currently not known precisely.  
We have therefore tested several  methods based on different assumptions regarding  $\alpha$, $RV_{\rm sppl}$, and $RV_{\rm conv}$ to estimate $\alpha$ from the {\it RV} time series themselves. 
Depending on the observation,
one assumption may be better than another. This approach should also allow us to  estimate the small-scale convection (such as granulation) amplitude in the star in addition to a corrected {\it RV}, for example as determined by \cite{meunier17}. 
Once $\alpha$ is estimated using one of these methods, eqs. 6 and 7 provide $RV_{\rm sppl}^r$, and $RV_{\rm conv}^r$.
The assumptions and methods are summarized in Table~\ref{tabmeth}.
\\

{\it Method 1.} 
We assume that $<RV_{\rm sppl}>=0$. This is not the case for $RV_{\rm conv}$, which is positive for all time steps. We note that in the reference series $RV_{\rm sppl}^t$=0.02~m/s and $RV_{\rm conv}^t$=3.17~m/s (Table.~\ref{tabref}). 
We search for the value of $\alpha$ that leads to a
reconstructed $RV_{\rm sppl}^r$ with an average of zero. 
The quality of the reconstruction when an offset is present is also  tested (Sect.~4.1);  although this 
assumption is correct for our simulated solar {\it RV}, this may not be the case for observed {\it RV}. \\

\begin{figure}
\includegraphics{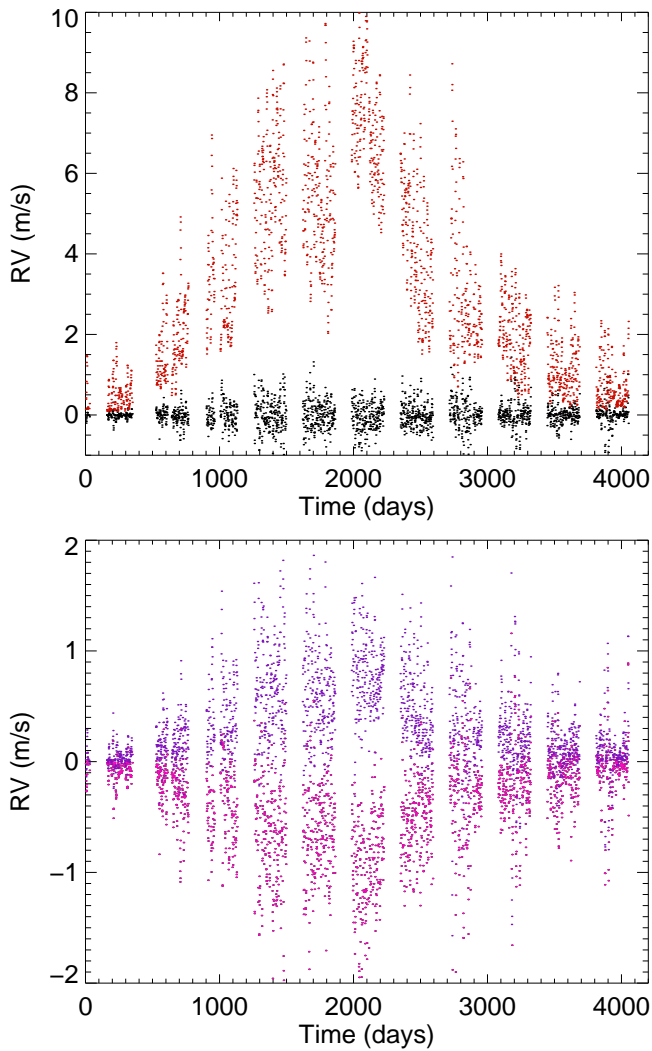}
\caption{
{\it Upper panel}: Reconstructed {\it RV} due to spots and plages $RV_{\rm sppl}^r$ (black) and convection attenuation in plages $RV_{\rm conv}^r$ (red), with no noise, for the set of lines S$_1$, full temporal sampling (with a 4 month gap every year), and method 1.
{\it Lower panel }: Reconstructed $RV_{\rm sppl}^r$ for a value of $\alpha$ that is 5\% too high (purple) and  5\% too low (pink) for the set of lines S$_1$, full temporal sampling (with a 4 month gap every year).
}
\label{nonoiseall}
\end{figure}

\begin{figure}
\includegraphics{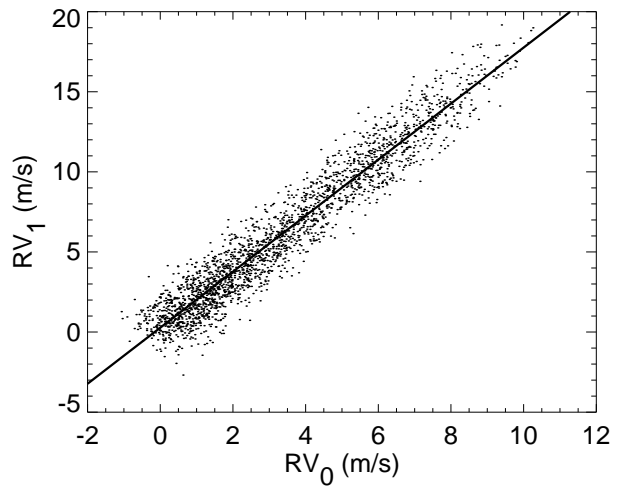}
\caption{
$RV_1$ versus $RV_0$ for a noise of 0.5 m/s (dots) and linear fit (solid line) for the sets of lines S$_1$ and S$_0$, respectively. }
\label{exrv}
\end{figure}

{\it Method 2.} 
We assume that the time series $RV_{\rm sppl}(t)$ and $RV_{\rm conv}(t)$ are uncorrelated. This is justified by the property of reference {\it RV} time series with a correlation between $RV_{\rm sppl}(t)$ and $RV_{\rm conv}(t)$ of 0.02, i.e., very close to zero. This is due to  the different natures of the {\it RV} signal  in the two cases: in the first case, the {\it RV} signal changes sign when the magnetic regions cross the central meridian \cite[e.g.,][]{desort07,lagrange10}. In the second case, the {\it RV} signal is always positive and reaches a maximum when the structures crosses the central meridian. 
We therefore determine the  unique value\footnote{ When $\alpha$ is not equal to the proper value, $RV_{\rm conv}$ contributes to the reconstructed $RV_{\rm sppl}^r$, and the correlation is then positive (resp. negative) if $\alpha$ leads to an underestimation (resp. overestimation) of $RV_{\rm conv}^r$ (see Fig.~\ref{nonoiseall}). }  of $\alpha$ that cancels the correlation between the reconstructed $RV_{\rm sppl}^r(t)$ and $RV_{\rm conv}^r(t)$. 

In the absence of noise, this technique gives a very precise value of $\alpha$. However, in the presence of noise,
this is not so and a correction must be performed. The reason is the following: the synthetic observed time series was built following eqs. 4 and 5.  
When deriving $RV_{\rm sppl}^r$ and $RV_{\rm conv}^r$ from $RV_0$ and $RV_1$
for a given $\alpha^r$,  these reconstructed time series depend on both $b_0$ and $b_1$. Therefore, the noise
in $RV_{\rm sppl}^r$ and $RV_{\rm conv}^r$  is correlated, leading to a shift in the correlation: in the presence of noise, instead of searching which value of $\alpha$ leads to a correlation of zero, we search for the value leading to the correlation due to the noise. 
We assume that the amplitude of the noise is well estimated for the set of lines considered. 
The amplitude of this effect is estimated and corrected for. \\

{\it Method 3. }
This method, as for the solar case, is based on the assumption that the convection signal dominates the total {\it RV} \cite{meunier10a}, and we use the relationship between $RV_1$ and $RV_0$. This can be checked on the reference series, especially during high activity periods, as $RV_{\rm sppl}^t$ has a rms  on the order of 0.3 m/s for an average close to 0, while $RV_{\rm conv}^t$ has a rms one order of magnitude larger and can reach values as high as 8-10 m/s as shown by \cite{meunier10a} and in Table~\ref{tabref}. 

In that case, the slope of $RV_1$ versus $RV_0$ is very close to $\alpha$. An example of $RV_1$ versus $RV_0$ is shown in Fig.~\ref{exrv} for a noise of 0.5 m/s, showing a slope of 0.67, while the true $\alpha$ is 0.70 (Table~\ref{tabmask}). We therefore perform a linear fit and derive an estimate of $\alpha$ from the slope.\\

{\it Method 4.}
This method is based on the same assumption as method 3, i.e., $RV_{\rm conv}$ amplitudes are much larger than $RV_{\rm sppl}$, but here we directly compare  the amplitudes of the {\it RV} signal.  
When $\alpha$ is properly determined, we expect the rms of $RV_{\rm sppl}^r$  to be small. If $\alpha$ is not properly determined, however, $RV_{\rm conv}$ can  leak into the reconstructed $RV_{\rm sppl}^r$, i.e., $RV_{\rm sppl}^r$ would include a fraction of $RV_{\rm conv}$ which may not be negligible with respect to $RV_{\rm sppl}$, which would increase its rms significantly. 
We minimize the ratio rms of $RV_{\rm sppl}$ / rms of $RV_{\rm conv}$.\\
 
{\it Method 5.}
This method is based on the same assumption as the previous method, but we consider long timescale variations. 
We minimize the rms of $RV_{\rm sppl}$ smoothed over 30 days. This method is the only one  sensitive only to long timescales, while the previous ones are sensitive to all timescales. The reason is that $RV_{\rm conv}$ presents some large-scale temporal variations (due to the solar cycle), while $RV_{\rm sppl}$ does not; therefore, the contribution of $RV_{\rm conv}$  to $RV_{\rm sppl}^r$ is easier to identify after removing the small-scale temporal variations.

\begin{table*}
\begin{center}
\caption{Assumptions and methods}
\label{tabmeth}
\begin{tabular}{cccc}
\hline \hline
Number & Assumption & Method & Timescales \\
\hline
1 & $<RV_{\rm sppl}>$=0  & $\alpha$ derived from this condition & all \\
2 & $RV_{\rm sppl}$(t) and $RV_{\rm conv}$(t) are uncorrelated  &  $\alpha$ derived from this condition &  all \\
3 & $RV_{\rm conv}$ dominates the signal  & $\alpha$ derived from the slope $RV_1$ versus $RV_0$ & all \\
4 & $RV_{\rm conv}$ dominates the signal     & minimization of the ratio rms $RV_{\rm sppl}$ / rms $RV_{\rm conv}$ & all \\
5 & $RV_{\rm conv}$ dominates the signal     & minimization of the rms of $RV_{\rm conv}$ over 30 days & long\\
 \hline
\end{tabular}
\end{center}
\end{table*}

\subsection{Step 3: Time series reconstruction characterization}

Because we know how the $RV_i$ series were built, we can compare the reconstructed $\alpha_i$ (case 2, for five methods), and the {\it RVs} with their true reference values.  We use three complementary criteria to compare the reference and reconstructed {\it RVs}: 
\begin{itemize}
\item{{\it The correlation between the reference and reconstructed time series.} A very good correlation indicates that the variations in the signal are well reproduced. We note that a correlation close to 1 may be obtained even if the proper amplitude is not retrieved, hence the following complementary criteria. }
\item{{\it The rms of the residuals between the reference and reconstructed series.} If the performance of the correction is good, this rms should follow the noise level. }
\item{{\it The correlation between $RV_{\rm sppl}^r$ and $RV_{\rm conv}^r$.} Although a small correlation is not sufficient to guarantee an excellent reconstruction at all timescales, a correlation different from zero means that the correction is not optimal and that the spot+plage residuals probably  contain a significant part of the convection signal.}
\end{itemize}

\subsection{Steps 4 and 5: {\it RV} correction and performance for exoplanet detectability}

Once we have obtained reconstructed times series, we correct $RV_0$ by subtracting the reconstructed $RV_{\rm conv}^r$. 

A straightforward estimation of the quality of the correction is obtained by directly comparing the {\it RV} time series. This is illustrated in Sect. 3.2. Given the number of simulations (for different S/N levels, methods, temporal samplings), it is   also necessary to quantify the quality of the correction using some criteria so that the methods can be compared and  the impact of the noise level on the performance can be studied more easily. We therefore use several
complementary criteria to characterize the residuals  (i.e., $RV_{\rm sppl}^r$):

\begin{itemize}
\item{The {\it rms {\it RV}} is computed and compared with the rms before correction and the best rms that can be theoretically achieved (i.e., the rms after correction with the reference $RV_{\rm sppl}^t$).}
\item{The periodogram of the corrected {\it RV} is computed and the {maximum power in four frequency domains} is derived: 2-10~d, 10-40~d (corresponding to the rotational period and harmonics), and 100-500~d,  500-800~d (both corresponding to long-term variability during the solar cycle), also to be compared with the power computed in the same ranges before correction (i.e., on $RV_0$) and on the time series after correction with the reference $RV_{\rm sppl}^t$. 
}
\item{The {\it detection limits} at 480~d (corresponding to 1.2 AU), as in \cite{lagrange10b} and \cite{meunier10a}, are computed using the local power amplitude (LPA) method \cite[][]{meunier12,meunier13} and compared with those before correction and after correction with the reference  $RV_{\rm sppl}^t$. We note that we use a revised version allowing a much faster computation, and with a slightly different threshold \cite[][]{lannier17}\footnote{The detectability criterion is that the maximum power due to the planet in the range 0.75-1.25$P_{\rm pla}$ (where $P_{\rm pla}$ is the planet period) is larger than 1.3 times the maximum power due to the observed signal. Computations are made for a fixed phase.
}.
}
\end{itemize}

We note that with an excellent correction (derived from an excellent estimation of $\alpha$), $RV_{\rm sppl}^r$ is the sum of mostly two components: the reference $RV_{\rm sppl}^t$ and some noise coming from both $b_0$ and $b_1$.

\section{Results}

\subsection{Parameters of the simulation}

In this section, we perform a simulation over all points covering one solar cycle using the properties described in Table~\ref{tabmask} for the set of line S$_1$, with S$_0$ used as a reference; we exclude a four-month period every year, as done in \cite{lagrange10} and \cite{meunier10a}, to simulate  that a given star is not observable at all times during the year, which introduces a one-year periodicity in the temporal sampling. 
The uncertainty on $\alpha$ for case 1 is chosen to be 5\%. 
This order of magnitude corresponds to the value obtained for noise below 0.5~m/s; therefore, it is an upper limit for a relatively good S/N (if an estimation of $\alpha$ in other conditions leads to a higher level, a scaling of the results must therefore be applied).
We first consider the case with no noise, then we consider different noise levels.
We note that although the case where $\alpha$ is known precisely is not realistic, it should give an upper limit to what can be done in an ideal case and allows an estimation of how close other cases are to this ideal situation. 

\subsection{No-noise case}

\begin{figure*}
\includegraphics{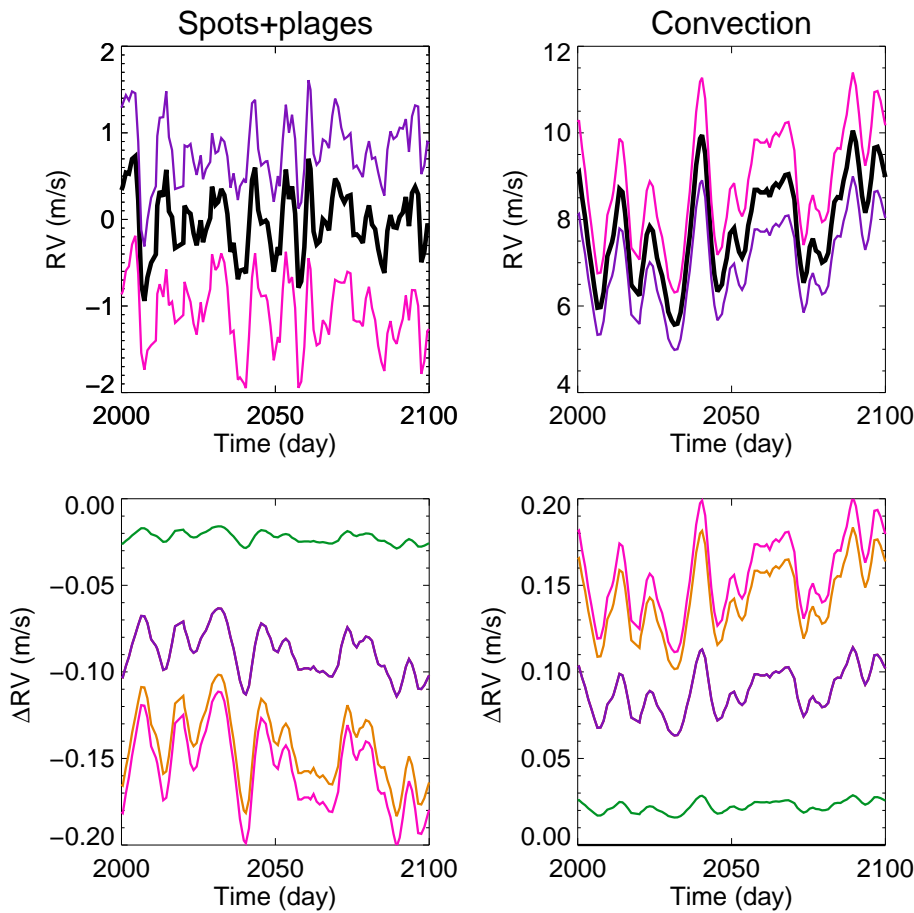}
\caption{
{\it Upper left panel}:  Reference $RV_{\rm sppl}^t$ (black thick line)  compared with the reconstructed $RV_{\rm sppl}^r$ for a value of $\alpha$ that is  5\%  too high (purple) and 5\% too low (pink line) for the set of lines S$_1$, full temporal sampling (with a 4 month gap every year) in the no-noise case.
{\it Upper right panel}: Same, but  for  $RV_{\rm conv}^r$.
{\it Lower left panel}: Reconstructed $RV_{\rm sppl}^r$ minus reference $RV_{\rm sppl}^t$ for the set of lines S$_1$, full temporal sampling (with a 4 month gap every year), no noise, and $\alpha$ fitted with different methods: method 1 (red, over which the purple curve is superimposed), method 2 (green), method 3 (orange), method 4 (pink), and method 5 (purple).
{\it Lower right panel}: Same, but  for  $RV_{\rm conv}^r$.
}
\label{nonoisesel}
\end{figure*}

We first consider the no-noise case. The different methods are explored and are compared with the case for which $\alpha$ is known precisely or with a given uncertainty.

Fig.~\ref{nonoiseall} (upper panel) shows the reconstructed {\it RV} for method 1 over the whole time range, which is representative of most methods used to fit $\alpha$. These reconstructed {\it RVs} can be compared to the reference  values shown in Fig.~\ref{rv}, and show a very good agreement. The lower panel of Fig.~\ref{nonoiseall} illustrates the impact of a bad estimation of $\alpha$: in this example ($\alpha$ over- or underestimated by 5\%), $RV_{\rm sppl}^r$ exhibits a long-term variation representing a fraction on the order of 10\% of $RV_{\rm conv}^t$ leading to an amplitude on the order of 0.8-1 m/s due to the error on $\alpha$. This illustrates the discussion for the choice of method 4 in Sect.~2.4.2.

Fig.~\ref{nonoisesel} shows a zoom on a limited time range during a high activity period for all cases and methods. The upper panels allow  the reconstructed {\it RVs} to be compared with the reference values for case 1, i.e., $\alpha$ known with a 5\% uncertainty, and for an exact value of $\alpha$. When $\alpha$ is exactly known, the reconstructed {\it RV} time series are exactly the same as the reference series. For $\alpha$ higher  or lower than the true value, however, the reconstructed values are offset by a significant amount, which is proportional to $RV_{\rm conv}^t$. As a consequence, $RV_{\rm conv}^r$, which can be used to correct the original signal for the convective contribution, differs from the true value by about 10 \%. This gives a good idea of the impact of the error of 5~\% on $\alpha$ on the quality of the reconstructed $RV_{\rm conv}^r$. 

The lower panels of Fig.~\ref{nonoisesel} shows the difference between the reconstructed and the reference time series in case 2, with $\alpha$ fitted using the five different methods. The  $RV_{\rm conv}^r$ time series differs from the reference values by  1.1 \% (methods 1, 2, 5), 1.8 \% (method 3), and 2 \% (method 4). The differences are slightly larger for $RV_{\rm sppl}^r$, with values between 1.9 and 3.4 \% depending on the method, and up to 18\% for the 5 \% error on $\alpha$ case.
We note that the difference is systematically negative for the spot+plage signal, and systematically positive for the convective component. This is due to the error on $\alpha$: as illustrated in Fig.~\ref{nonoiseall}, the sign of the error on $\alpha$ controls the sign of the difference between reconstructed and true value. 

In the absence of noise, we therefore obtain excellent reconstructed {\it RV} time series, which should allow us not only to correct properly for the convective contribution to {\it RV}, but also to study very precisely the {\it RV} variations due to activity themselves.

\subsection{Impact of noise}

\begin{figure*}
\includegraphics{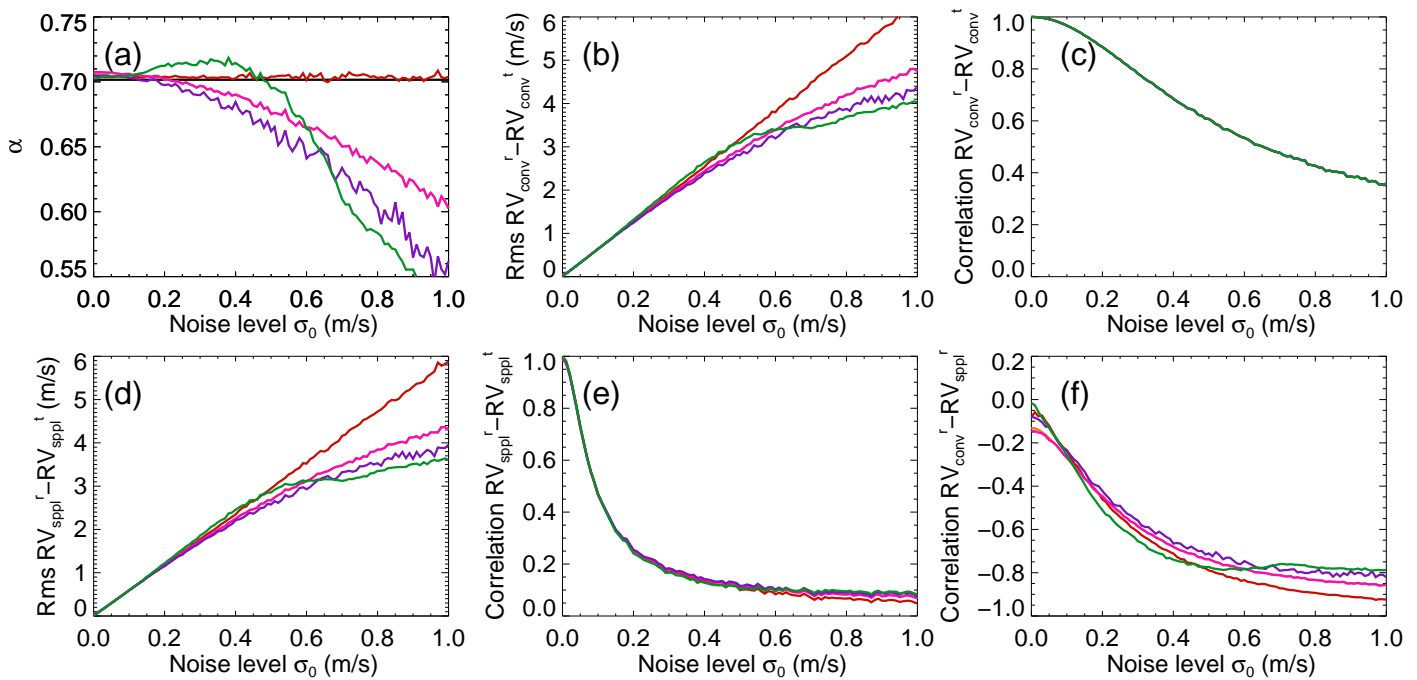}
\caption{
{\it Panel (a)}: Reconstructed $\alpha$ versus $\sigma_0$ for S$_1$, full temporal sampling (with a 4 month gap every year) and different methods (see Fig.~6, lower panels, for the color-coding; the curves for methods 3 and 4 are almost indistinguishable here).
All noise realizations have been averaged.
The true value is indicated by a solid line (only for this panel).
{\it Panel (b)}: Same, but for the rms {\it RV} of $RV_{\rm conv}^t-RV_{\rm conv}^r$.
{\it Panel (c)}: Same, but  for the correlation between $RV_{\rm conv}^t$ and $RV_{\rm conv}^r$.
{\it Panel (d)}: Same, but  for the rms {\it RV} of $RV_{\rm sppl}^t-RV_{\rm sppl}^r$
{\it Panel (e)}: Same, but  for the correlation between $RV_{\rm sppl}^t$ and $RV_{\rm sppl}^r$.
{\it Panel (f)}: Same, but  for the correlation between $RV_{\rm conv}^r$ and $RV_{\rm sppl}^r$.
}
\label{set1a}
\end{figure*}

\begin{figure}
\includegraphics{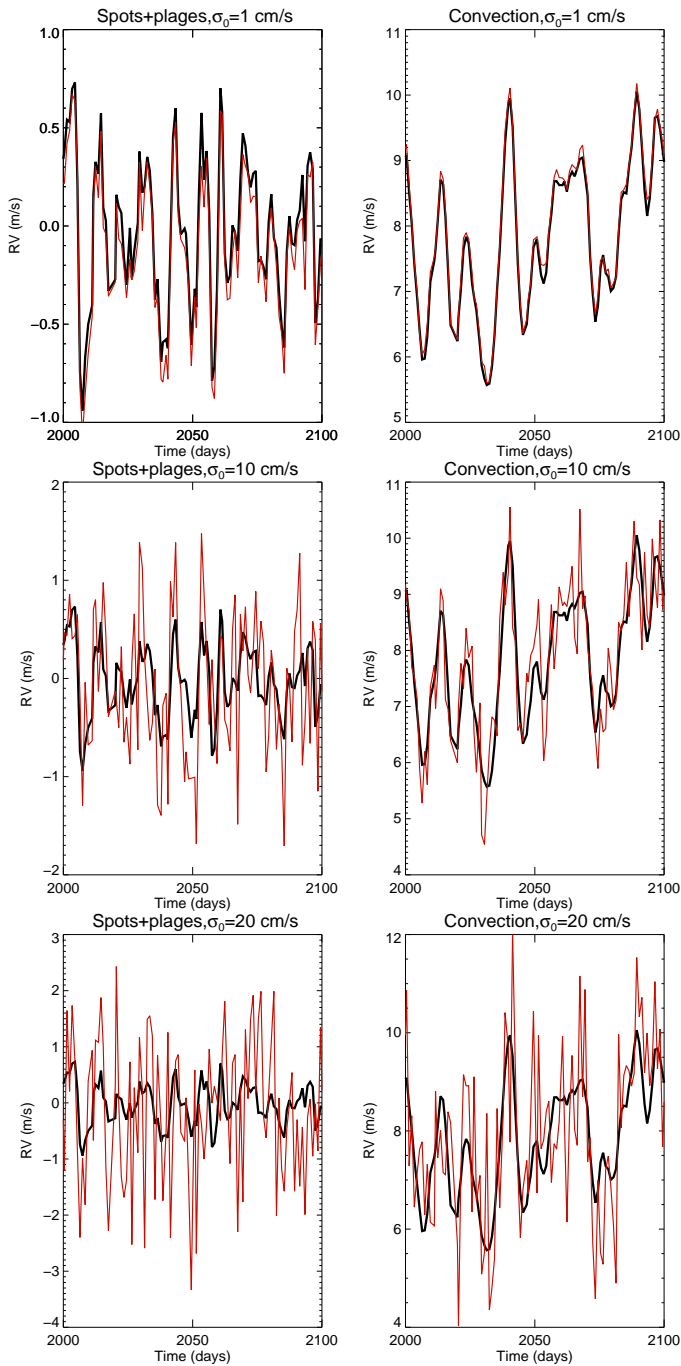}
\caption{
{\it Upper panels}: Reference {\it RV} component (black) and reconstructed (red) {\it RV} computed with method 1 during a period of high activity for  the set of lines S$_1$, for the spot + plage (left columns) and convection (right columns), for $\sigma_0$=1~cm/s. 
{\it Middle panels}: Same, but  for $\sigma_0$=10~cm/s.
{\it Lower panels}: Same, but  for $\sigma_0$=20~cm/s. 
}
\label{ex_serie}
\end{figure}

\begin{figure*}
\includegraphics{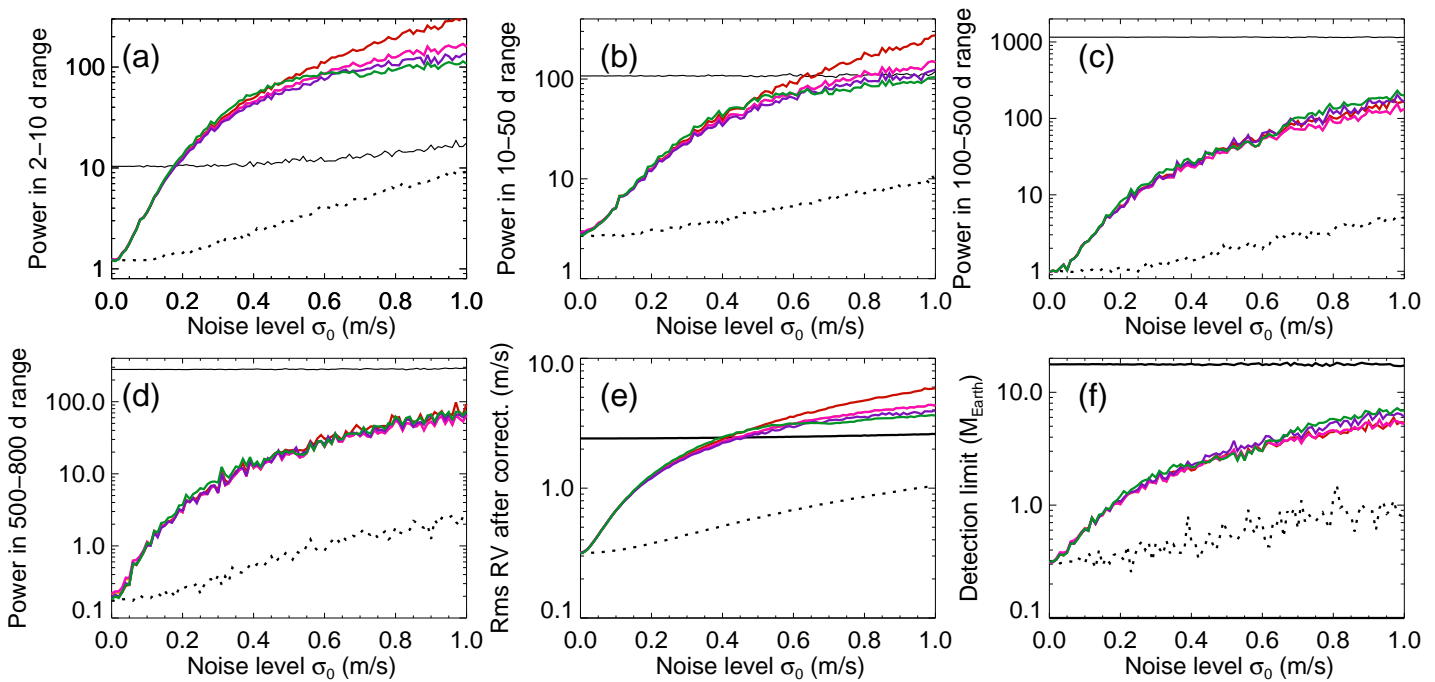}
\caption{
{\it Panel (a)}: Maximum power in the 2-10~d range computed on the periodogram of the {\it RV} residuals after correction versus $\sigma_0$ for S$_1$, full temporal sampling (with a 4 month gap every year) and different methods (see Fig.~6, lower panels, for the color-coding; the  curves for methods 3 and 4 are almost indistinguishable here). The solid black line shows the power before correction and the dotted black line the power after correction in an ideal case (i.e., correction with the reference $RV_{\rm conv}^t$).
{\it Panel (b)}: Same, but  for the power in the range 10-50~d.
{\it Panel (c)}: Same, but  for the power in the range 100-500~d.
{\it Panel (d)}: Same, but  for the power in the range 500-800~d.
{\it Panel (e)}: Same, but for the rms of the residuals after correction.
{\it Panel (f)}: Same, but  for the detection limits at 480~d.
}
\label{set1b}
\end{figure*}

\begin{figure}
\includegraphics{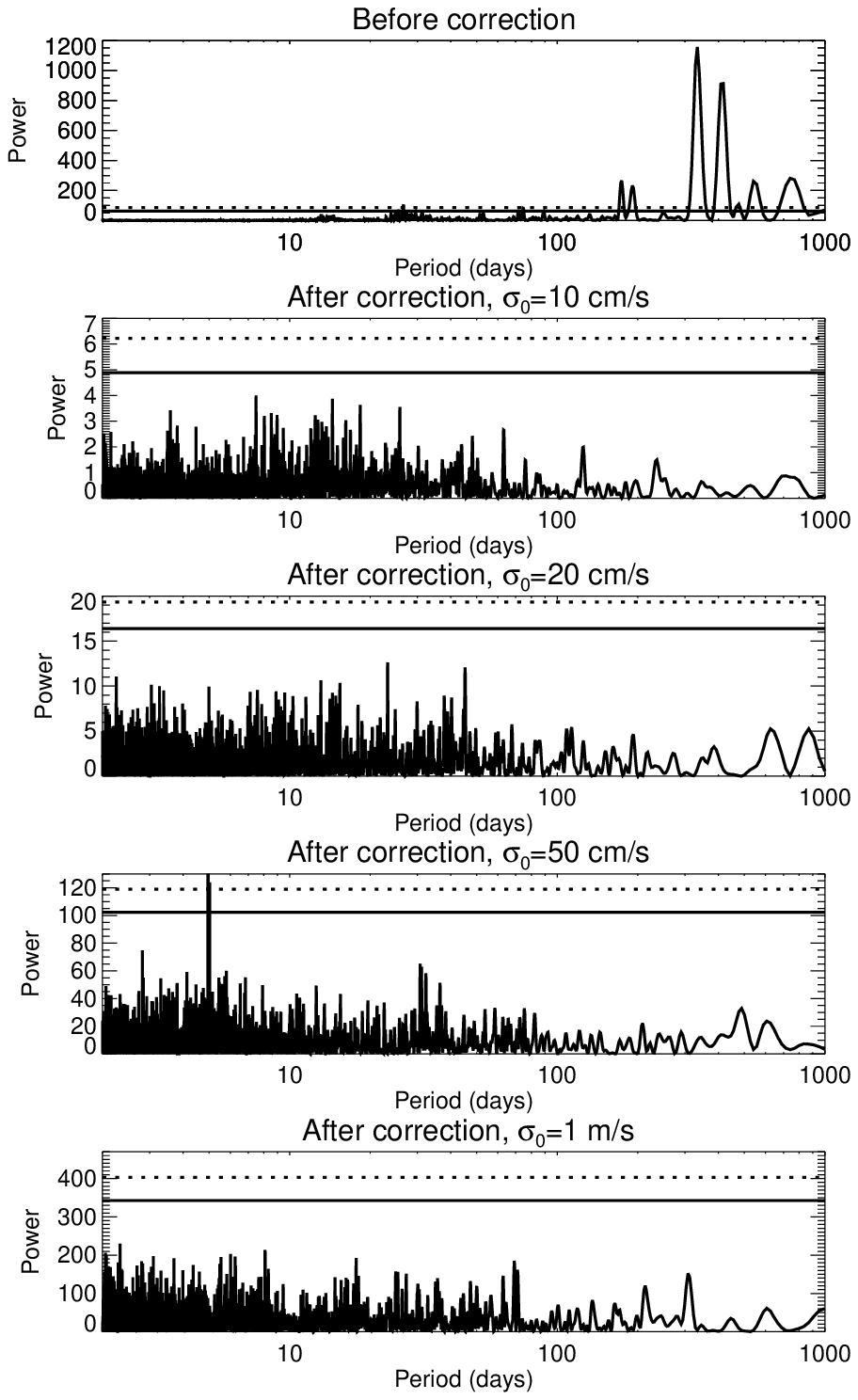}
\caption{
{\it First panel}: Periodogram of the simulated time series before correction (all points except for a 4 month gap), for a $\sigma_0$  of 10~cm/s. 
{\it From second to fifth panel}: Same after correction using method 1 and the set of lines S$_1$, respectively for a $\sigma_0$  of 10~cm/s, 20~cm/s, 50~cm/s, and 1~m/s. The horizontal lines show the false alarm probability (fap) at 1~\% (dashed lines) and 10~\% (solid lines).
}
\label{periodo}
\end{figure}

\subsubsection{Validation of the reconstructed series}

We first compare the reconstructed $\alpha$ with the true values. 
The results are shown in Fig.~\ref{set1a} (panel a) for various $\sigma_0$  and methods.
For low noise levels (below 20~cm/s), the reconstructed $\alpha$ is very good.
The reconstructed $\alpha$ remains within 5\% of the true value up to 50~cm/s. For higher noise levels (up to 1~m/s), 
method 1 (within perfect conditions), always leads to good results and is therefore quite insensitive to noise. The other methods are all divergent, however. 

We now compare the reconstructed {\it RV} with the reference value using the correlation 
between reference and reconstructed {\it RVs} and the rms {\it RV} of the difference in Fig.~\ref{set1a} (panels b to f).
Let us consider first the reconstruction of the convective component $RV_{\rm conv}^r$. The rms of the difference with the reference {\it RV} series (panels b and c) 
naturally increases with noise, reaching $\sim$1~m/s for a noise level around 20~cm/s. This is observed for all methods. 
The correlation between $RV_{\rm conv}^r$ and $RV_{\rm conv}^t$ (panel c) decreases as the noise increases, reaching values of 0.8 around 25~cm/s for all methods, and 0.4 for a noise above 80~cm/s.

As for $RV_{\rm sppl}^r$, which is of great interest because it is the residual after correction of the convection signal, the rms of the difference with the reference {\it RV} series (panel d) are globally similar to the convective component. 
The correlation (panel e)  on the other hand decreases towards 0 much faster, showing that even for low noise levels it is impossible to reproduce the temporal variation of this component in a realistic way. 
Only for a noise level of a few cm/s would this be possible.

Finally, the correlation between $RV_{\rm sppl}^r$ and $RV_{\rm conv}^r$ is shown in panel f. In principle this correlation should be close to zero. If it is not the case, it means that $RV_{\rm sppl}^r$ includes a significant part of the convective signal as the large amplitude of the latter dominates the correlation. This correlation is close to zero for a noise level of just a few cm/s.

Fig.~\ref{ex_serie} shows an example of reconstructed time series with method 1 during a period of high activity for three different  noise levels (1, 10, and 20 cm/s). $RV_{\rm sppl}^r$ is noisier than $RV_{\rm conv}^r$. It is possible to recognize some short-term variations, although it is noisier than the reference signal, only for very low noise levels (cm/s). The convective signal is better reproduced up to 10~cm/s.
Naturally, the very good agreement for the convective contribution is crucial because it shows that it is reasonably possible to  correct for it in good conditions.

\subsubsection{Performance for exoplanet detectability}

We characterize the {\it RV} residuals after the correction with $RV_{\rm conv}^r$ by computing their rms, the power of the periogram in various ranges, and detection limits at 480 days,  as described in Sect.~2.6.
These detection limits can also be compared to those found by \cite{meunier13} using a correction based on the calcium index (hereafter Ca correction). The Ca correction used in this paper represents the chromospheric emission, which is directly related to the surface covered by plages and therefore also directly related to $RV_{\rm conv}^t$. This is a variable that can be determined from stellar observations.

The  maximum  power in four period ranges is shown in Fig.~\ref{set1b} (panels a to d), illustrating synthetically how the periodograms evolve with noise before and after correction. 
The power is always increasing with $\sigma_0$, all methods performing similarly. The gain in power is the best for the power in the range 100-500~d (of great interest for Earth-like planets in the habitable zone around solar-type stars) and 500-800~d, for which the gain can reach three orders of magnitude at very low noise levels. The gain is about one order of magnitude only around 0.6 m/s for the 100-500~d range. On the other hand, for the power at low periods, the gain is much smaller and a significant gain is achieved only for low noise levels: the power is the 2-10 range is {higher} than before correction for $\sigma_0$ above 20~cm/s, and in the 10-40~d range it reaches the power before correction around 60~cm/s.  When performing the correction, we therefore add a significant amount of noise at high frequencies. 

Fig.~\ref{periodo} shows a few examples of periodograms (1 out of the 20 realizations) before and after correction (only one plot is shown before correction as they are very similar for the different noise levels). The periodogram before correction shows some strong peaks in the period range of  100-800: this strong power has already been noticed by \cite{meunier10a} and is due to variations in the filling factor of plages (and network) during the solar cycle. Fig.~\ref{periodo}  illustrates how well the power is reduced at all frequencies for a very low noise level ($\sigma_0$ of 10~cm/s), with power and false alarm probabilities (fap) much lower than before correction. 
The 1~m/s plot exhibits a much higher power after correction, which is comparable to the power at long periods obtained when using the Ca correction for a medium Ca noise level \cite[see Fig.~17 in][]{meunier13}, although there is much more noise here at low periods. However, the power in the range of 100-500 days obtained for a $\sigma_0$ of 50 cm/s is  better than that  obtained with the medium Ca noise level.  The typical faps are lower than the fap before correction for $\sigma_0$  below 40~cm/s, as is the maximum power: this is similar to what is  obtained below when comparing the rms {\it RV} before and after correction. Finally, we note that for the example shown for 50~cm/s there are a few high peaks at periods around a few days and around 30 days: these peaks are not present for the other realizations. At the different noise levels, there are indeed a few realizations for which we do observe such peaks, most of the time below the 1~\% and 10~\% fap, but there a few cases for which they are above them. We note that for $\sigma_0$  below 10~cm/s the maximum power is above the fap but corresponds to a true power (rotation modulation). We have quantified the number of such peaks as a function of noise outside the rotational modulation period range, and found that the power is higher than the 1\% fap level in one realization at most.

The rms of the residuals are shown in panel e in Fig.~\ref{set1b} and the detection limits in panel f. The rms remains below 1 m/s for $\sigma_0$  lower than 15~cm/s, but is above the rms {\it RV} before correction above 40~cm/s. 
The detection limits are very low at low $\sigma_0$: they are below 1 M$_{\rm Earth}$ for $\sigma_0$  lower than 15~cm/s. For $\sigma_0$ lower than 10~cm/s, they are also below the value of 0.8 M$_{\rm Earth}$ found for the Ca correction with high Ca S/N in \cite{meunier13} for most methods.  For the largest $\sigma_0$, the detection limit may be better than before correction (and could correspond to the super-Earth regime), while the correction does increase the rms of residuals: this larger rms is due mostly to an increase in power at small timescales, and in these cases the correction is  to be taken with caution despite the gain in detection limit.

\begin{figure}
\includegraphics{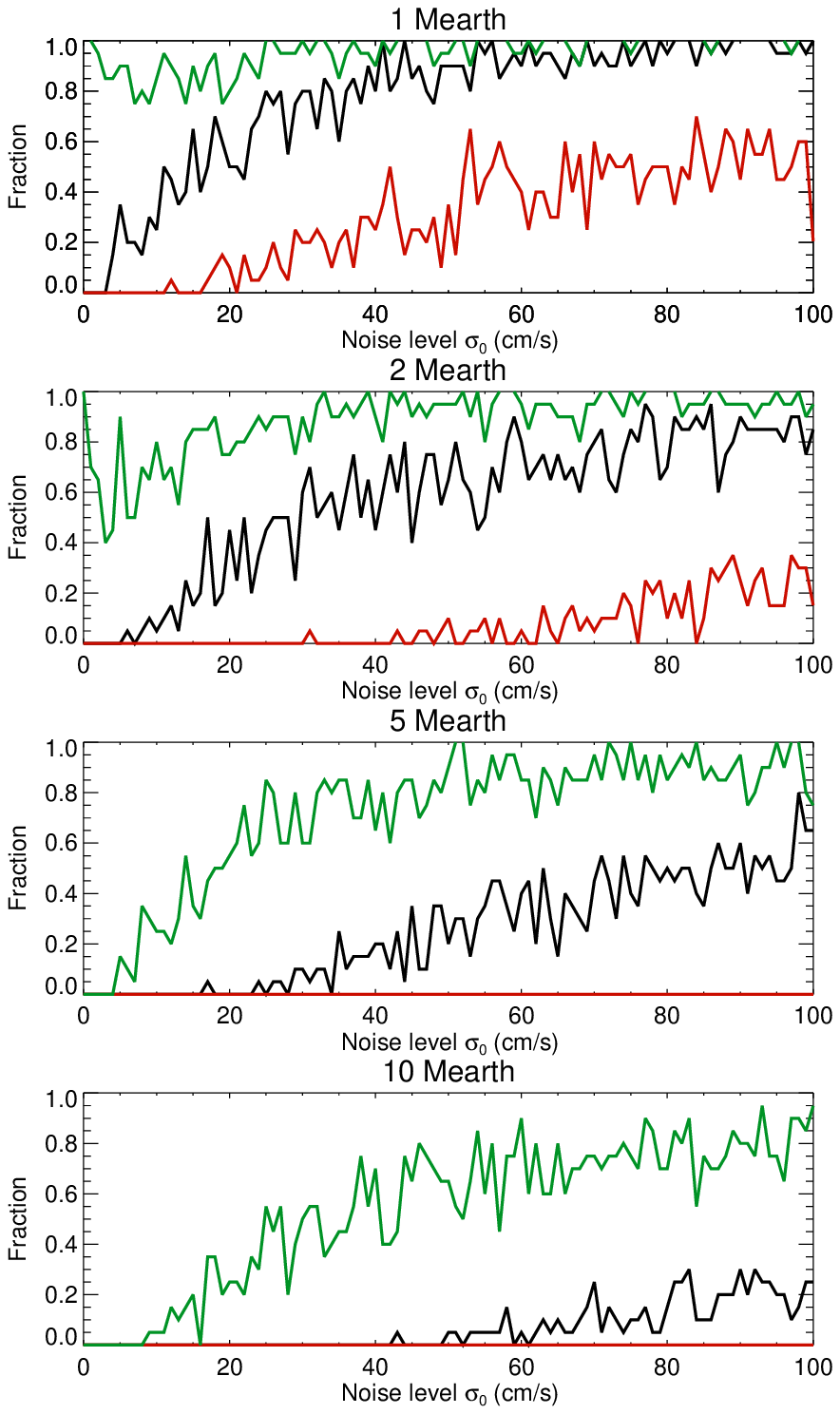}
\caption{
{\it First panel}: Fraction of realizations for which the planet amplitude after correction differs by more than 50\% (black) and 10\% (green) from the theoretical value. The red curve shows the 50~\% curve for the signal $RV_{\rm sppl}^t$+planet+noise, i.e., what would be obtained with a perfect correction.
{\it Second panel}: Same, but  for 2~M$_{\rm Earth}$.
{\it Third panel}: Same, but for 5~M$_{\rm Earth}$ (dotted line at the zero level).
{\it Fourth panel}: Same, but  for 10~M$_{\rm Earth}$ (dotted line at the zero level). 
} 
\label{ajoutpla}
\end{figure}

Finally, we performed an additional test adding a planetary signal (planet with masses 1, 2, 5, and 10 M$_{\rm Earth}$) at the same period (480~d) before applying our correction methods. Our objective is to see how the peak corresponding to the planet behaves as the noise level increases in order to check whether the correction impacts that peak. The amplitude of the peak (i.e., the power at these periods) in the periodograms for these planets alone is around 4, 16, 100, and 400, respectively, which can be compared to the power in Fig.~\ref{periodo}. At $\sigma_0$=10~cm/s, the planetary peak remains mostly unaffected for the four tested masses because of the low noise level. However, for higher noise levels, the number of realizations for which the planet peak amplitude is modified increases. This is illustrated in Fig.~\ref{ajoutpla}: the solid line shows the fraction of realizations for which the planet peak amplitude after correction differs from the expected value by more than 50\% for the four planet masses. For 1 and 2 M$_{\rm Earth}$, this fraction represents more than half the realizations for $\sigma_0$  above 20~cm/s and 30 cm/s, respectively, although this fraction is much lower for larger masses. The threshold of 10~\% is represented by the dashed lines: even for 10 M$_{\rm Earth}$, more than half the realizations lead to a difference of more than 10~\%. Finally, we also show the same fraction (for the 50\% threshold) computed for $RV_{\rm sppl}^t$+planet+noise, i.e., what would be obtained with a perfect correction: we also observe a significant impact on the planet peak amplitude, but smaller than the impact after correction, showing that a significant part of the variation is related to the correction.
Care should therefore be taken when interpreting the planet peak amplitudes.


\section{Discussion of our assumptions}

\subsection{Impact of assumptions in the different methods}

Method 1 is very promising. However, it relies on a strong assumption: the signal is the addition of $RV_{\rm sppl}$ with a zero average and of $RV_{\rm conv}$. On real observations the true zero of {\it RVs} is not necessarily known with a good precision. If an offset is added to the simulated signal the assumption is no longer true,  and this indeed leads to a bias. We have tested the impact of this issue by adding an offset of 2 m/s to the simulated {\it RV}. This choice is arbitrary, but given the typical {\it RV} variations such as those in Fig.~\ref{rv}, we estimate that if the convection inhibition is important, it should be possible to estimate the {\it RV} zero 
within this uncertainty or possibly better. This is a realistic value given the average of the total signal, although for a well-observed star it may be lower. A strong bias is observed, even with no noise: instead of a value close to 0.70 we find $\alpha$=0.81, which is significantly outside the $\pm$5\% range and correspond to typical biases obtained for $\sigma_0$ above 0.8~m/s for the other methods. The gain in power is very small in all period ranges, even for a very good S/N, and the detection limits  remain above 10~M$_{\rm Earth}$.

In methods 3 to 5, we assume that the convection signal is much larger than the spot+plage signal. While this is true for the Sun, it may not be true for other stars. We therefore performed a similar simulation with the convective signal divided by a factor of two, so that the relative amplitude between $RV_{\rm conv}$ and $RV_{\rm sppl}$ is smaller (ratio divided by a factor two). Method 1  performs similarly to the previous case, but the other methods  all diverge faster from the true $\alpha$ value as the noise increases, reaching a 3\% difference around 10 cm/s. Methods 3 and 4 also show a bias on that order of magnitude even when no noise is present. However, the rms {\it RV} between the reconstructed and reference {\it RV} series are similar up to 30~cm/s and then much better (except for method 1, no decrease) than for the full convection signal for higher noise levels: although $\alpha$ is more poorly reconstructed, the correction performs well.

\subsection{Impact of the temporal sampling}

\begin{figure*}
\includegraphics{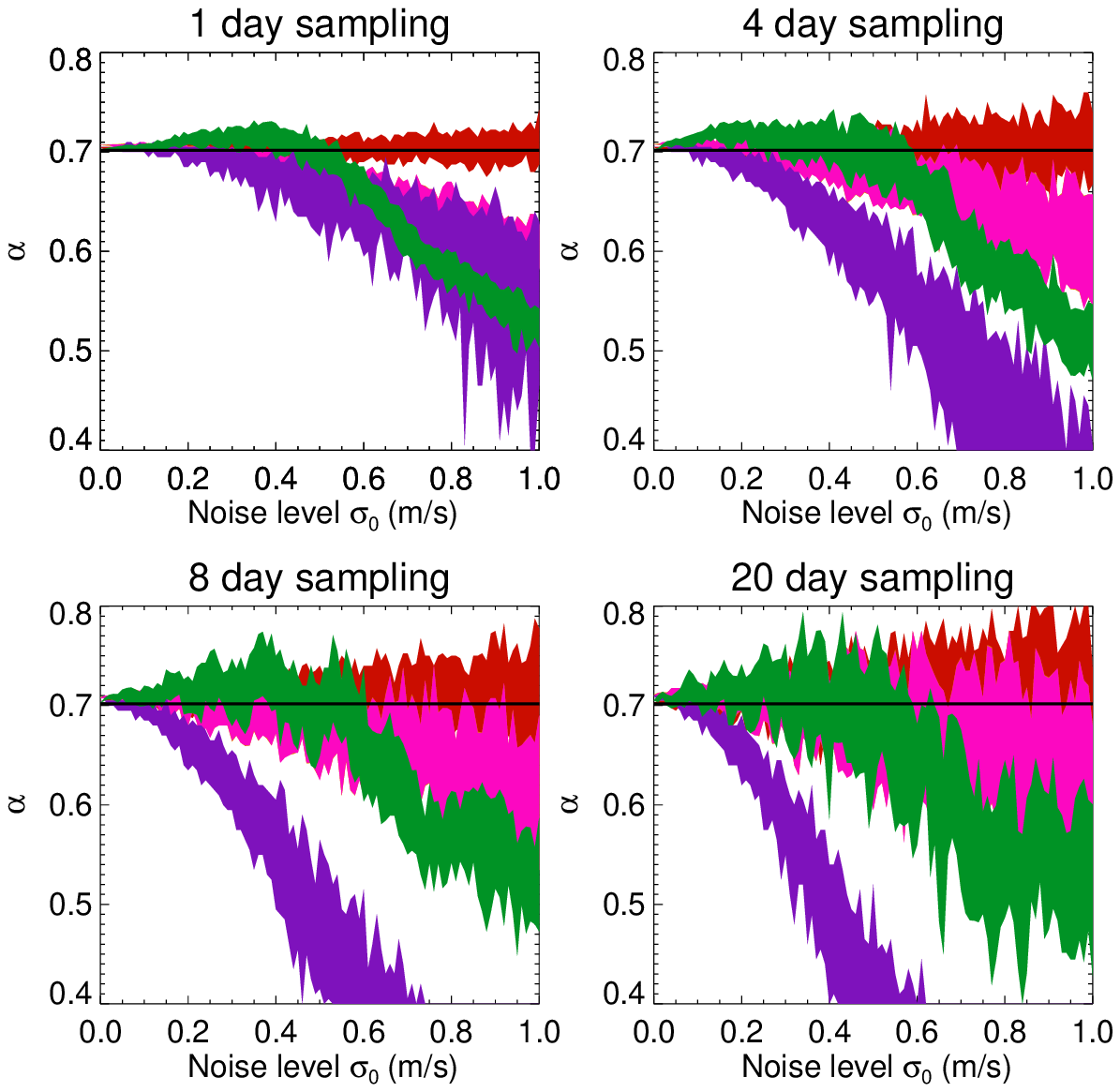}
\caption{
{\it Upper left panel}: Reconstructed $\alpha$ versus the noise level for the set of lines S$_1$ and all realizations for a sampling of 1 day (see Fig.~5 for color-coding).
{\it Upper right panel}: Same, but  for 4 days.
{\it Lower left panel}: Same, but  for 8 days.
{\it Lower right panel}: Same, but for 20 days.
}
\label{alpha_samp}
\end{figure*}

\begin{figure}
\includegraphics{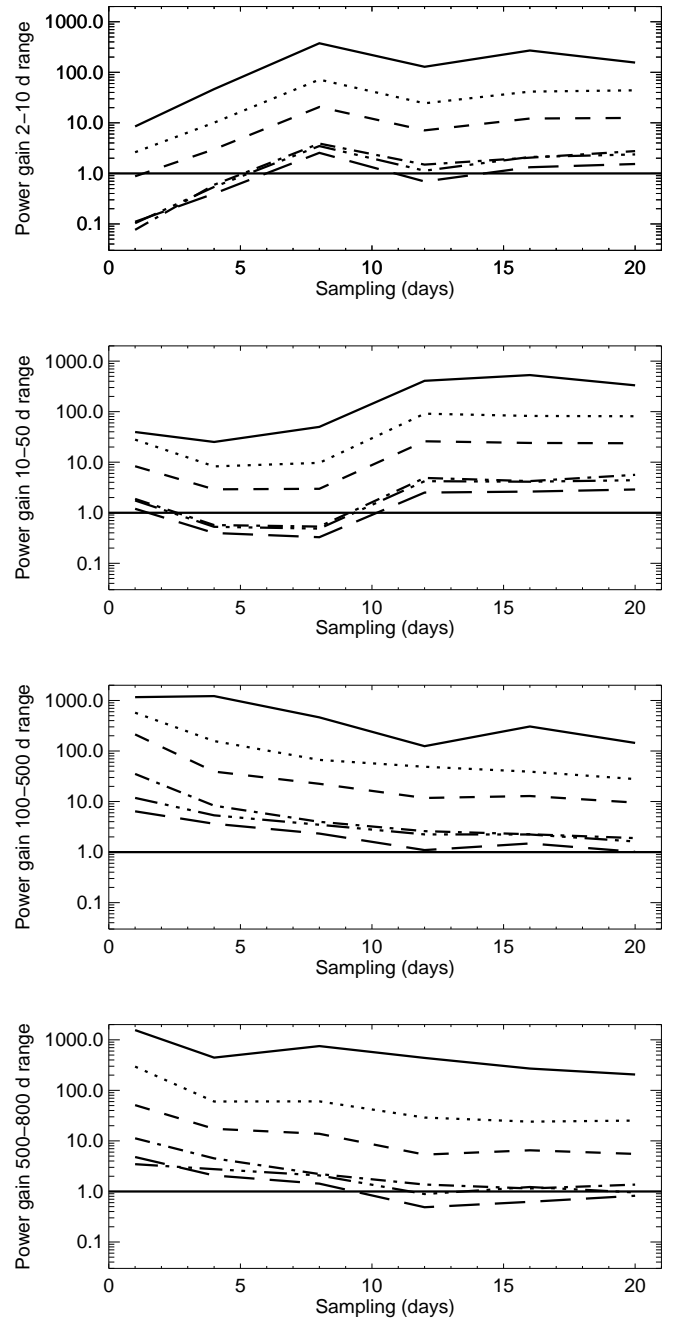}
\caption{
{\it First panel}: Ratio between the maximum power (in the range 2-10~d) in the periodograms before correction and  after correction (method 1) for various noise levels, showing the gain in power: no noise (solid line), 10~cm/s (dotted line), 20~cm/s (dashed line), 50~cm/s (dot-dashed line), 75~cm/s (dot-dot-dot-dashed line), 1 m/s (long-dashed line). The horizontal solid line represent a gain of A (i.e., no improvement).
{\it Second panel}: Same, but  for the period range 10-40~d.
{\it Third panel}: Same, but  for the period range 100-500~d.
{\it Fourth panel}: Same, but for the period range 500-800~d.
}
\label{ratio_power}
\end{figure}

\begin{figure}
\includegraphics{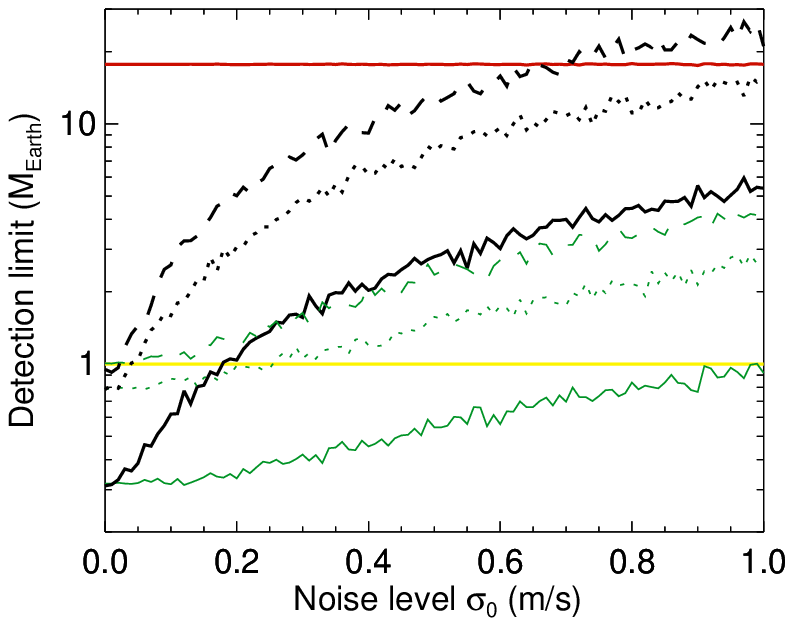}
\caption{
Detection limits versus the noise levels after correction for various sampling (black lines): 1 day (solid line), 8 days (dotted line), and 20 days (dashed line), averaged over the 20 realizations of the noise. The green curves show the detection limit for $RV_{\rm sppl}^t$ for the same sampling (same line code). The upper horizontal red line corresponds to the detection limit before correction for the 1 day sampling, and the horizontal yellow line the 1~M$_{\rm Earth}$ detection limit level. 
}
\label{limdet_samp}
\end{figure}

 We consider now the same sampling as in our previous works, i.e., we select one point every 4, 8, and 20 days in our time series including a four-month gap, covering the full 12.5 year duration to which we have added 12 and 16 day samplings.

The estimated $\alpha$ for S$_1$ are shown in Fig.~\ref{alpha_samp} for samplings of 4, 8, and 20 days and compared to the 1 day sampling. The trends are similar, but the estimation of $\alpha$ gets noisier as the sampling is degraded. Method 5  diverges much faster than the other methods as the sampling is degraded.  For the 4 day sampling, $\alpha$ remains with 3\% of the true value for  noise below 10-15~cm/s (instead of 20-25~cm/s for the 1 day sampling), 10~cm/s for a sampling of 20 d. 

Fig.~\ref{ratio_power} shows the gain in terms of maximum power for different period ranges after correction for the different temporal samplings and different noise levels for method 1. 
For the power in range 100-500 and 500-800 d, the gain is usually larger than 1 except for  high noise levels and for highly degraded sampling, and decreases as the sampling is degraded. On the other hand, the gain increases at lower periods as the sampling is degraded, and reaches 80-100 for very good noise levels and degraded sampling, while it is around 40 for  good sampling.

Finally, the detection limits increase as the temporal sampling is degraded for all noise levels, as shown in Fig.~\ref{limdet_samp}. While for all points a 1 M$_{\rm Earth}$ detection limit was obtained for noise below $\sim$ 10~cm/s, this threshold falls to $\sim$ 8~cm/s for the 4 day sampling and to  $\sim$ 4~cm/s for the 8 day sampling. It is only marginally lower than 1 M$_{\rm Earth}$ for the 20 day sampling (no noise). This is not due to the correction performance, however, as a perfect removal of the convection signal leads to a detection limit close to 1 M$_{\rm Earth}$ or above due to the spot+plage signal as well, as shown by the lower limit (green curves).

\subsection{Impact of other sources of noise}

\begin{figure}
\includegraphics{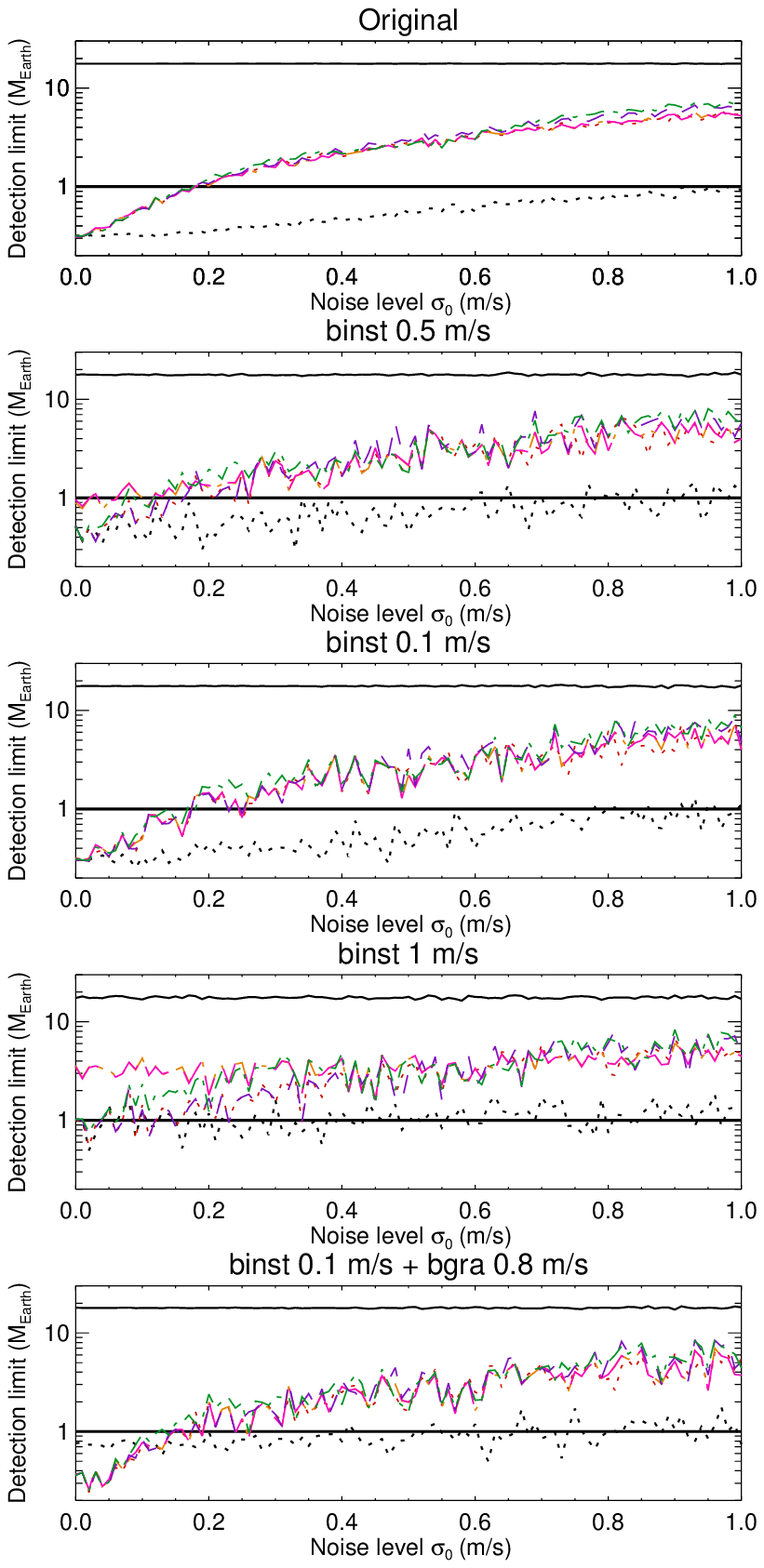}
\caption{
{\it First panel}: Detection limits versus noise for the full sample and no additional noise, averaged over all realizations. Color- and line-coding as in Fig.~13, panel f. The horizontal line is the 1~M$_{\rm Earth}$ detection limit level.  
{\it Second panel}: Same, but for $b_{\rm inst}(t)$ of 0.5 m/s, for one realization.
{\it Third panel}: Same, but for $b_{\rm inst}(t)$ of 0.1 m/s, for one realization.
{\it Fourth panel}: Same, but for $b_{\rm inst}(t)$ of 1 m/s, for one realization.
{\it Fifth panel}: Same, but for $b_{\rm inst}(t)$ of 0.1 m/s added to $b_{\rm gra}(t)$ of 0.8~m/s, for one realization.
}
\label{limdet_noise}
\end{figure}

In this work, we  considered only the {\it RV} noise due to the {\it RV} computation on noisy spectra. This noise depends on the chosen set of lines. In this section we study the impact of two other types of noise with different properties and test the impact of these contributions for all points except for the four-month gap every year (to be compared with the results shown in Sect.~3):
\begin{itemize}
\item{Instrumental instability: this contribution is independent of the set of lines and is exactly the same for all time series. A contribution $b_{\rm inst}(t)$ should therefore be added in eqs. 4 and 5. In this work we consider a contribution of 1 m/s (corresponding to current instrumental HARPS performance), 10~cm/s \cite[corresponding to future instruments, e.g.,][]{dodorico07}, and 50~cm/s (intermediate amplitude).
}
\item{The {\it RV} noise at high frequency due to convection, and in particular granulation, should also be considered. This noise is due to the stochastic realization of many granules covering the surface at a given time. It varies from one observation to the next. This should not be confused with the convective inhibition due to magnetic fields, which is the main subject of this paper, and which varies on much longer timescales (we  therefore use the term granulation in the following).} We use the granulation {\it RV} time series  derived by \cite{meunier15} in the solar case for a whole cycle (this signal is due to the different realizations of granules on the surface at each time step). The signal $b_{\rm gra}$ is added to $RV_0$ (eq. 4). As for $RV_i$ (eq. 5), we add the same time series, but modulated in amplitude because the amplitude of the granule velocities depends on the spectral lines, which controls both $RV_{\rm conv}$ and $b_{\rm gra}$. We make the assumption that the factor is similar to the factor controlling $\Delta V$ and therefore add $\alpha b_{\rm gra}$ in eq. 5. This means that $b_{\rm gra}$ is corrected at the same time as $RV_{\rm conv}$.
\end{itemize}

We first consider $b_{\rm inst}(t)$ only, with an amplitude of 0.5~m/s. It is added to the noise $b_i$ already considered in this paper. If $\alpha$ is exact, then $RV_{\rm conv}^r$ is the same as before because $b_{\rm inst}(t)$ is the same in $RV_0$ and $RV_1$, and therefore does not impact eq. 6. $RV_{\rm sppl}^r$  on the other hand includes that additional noise. More generally, because there is  additional noise on $RV_0$ and $RV_1$, the estimation of $\alpha$ is not as good as before: the very small bias at very low noise levels observed for methods 3 and 4 (Fig.~\ref{set1a}, panel a) is amplified to 3\% (e.g.) as it is for method 2. The global trend of the reconstructed $\alpha$ remains very similar, however. The rms between the reference and reconstructed values is slightly larger but this is probably a direct consequence of the additional noise.  
The detection limits illustrated in Fig.~\ref{limdet_noise} are not as good, however, with minimal values for the no-noise case around 0.5~M$_{\rm Earth}$ for methods 1, 2, and 5 and close to 1~M$_{\rm Earth}$ for methods 3 and 4, while they were all around 0.3~M$_{\rm Earth}$ without $b_{\rm inst}$. As a consequence they are lower than 1~M$_{\rm Earth}$ for very small noise levels only.  
For $b_{\rm inst}$ with an amplitude of 1~m/s, the biases on $\alpha$ for methods 3 and 4 reaches 7\% and method 2 diverges faster as well. Detection limits of 1 M$_{\rm Earth}$ are only marginally achievable when there is no noise on the {\it RV} determination (apart from the $b_{\rm inst}$ contribution).
For $b_{\rm inst}$ with an amplitude of 0.1~m/s, however, which should be reachable with future instruments, the results are very similar to the case with no instrumental noise as this contribution does not impact our results.

We now consider the second contribution, granulation. We add this contribution \cite[which has a rms {\it RV} of 0.8~m/s, from][]{meunier15} to an instrumental noise $b_{\rm inst}(t)$ of amplitude of 0.1~m/s that can be expected from future instruments. 
We find that the reconstructed $\alpha$ is very close to the value obtained without these additional  noise contributions. The same is observed for the correlation and rms of the differences between the reconstructed and reference time series. 
The power also performs very well. For example, the power in the range 100-500~d presents a gain greater than 100 for noise below 25~cm/s. For very low noise levels (i.e., with contribution from $b_{\rm inst}$ and $b_{\rm gra}$ only), the gain is almost 3 orders of magnitude. The impact on the correction of  low level of  instrumental noise and a realistic granulation time series is therefore very small. The detection limits, shown in the lower panel in Fig.~\ref{limdet_noise}, are below 1 M$_{\rm Earth}$ for a range of noise similar to the no-granulation case, i.e., below 10~cm/s. It should be noted, however, that these detection limits are lower than those obtained with a correction with the reference $RV_{\rm conv}^t$. This is due to the fact that $b_{\rm gra}$ behaves as $RV_{\rm conv}$, i.e., contributes with a factor $\alpha$ to $RV_1$. It is therefore included in the correction made with our method, hence a small impact on our detection limits. On the other hand, after correction of the reference $RV_{\rm conv}$ only, and due to the presence of $b_{\rm gra}$, the power is greater even at large periods, leading to a slightly higher detection limit.

%
%

\subsection{Note on application of the method to current HARPS data}

Our method leads to very good results; the  detection limits are around 1~M$_{\rm Earth}$ for very low noise levels, typically for $\sigma_0$ lower than 10~cm/s. It is therefore difficult to apply to current HARPS data since the noise level is much higher than this. Most of the time the sampling is not as good either. In principle, a very high S/N on the spectra could be compensated by temporal averaging, however, and detection limits of a few M$_{\rm Earth}$ could be obtained for higher noise levels. 

We test our method on a time series of 257 spectra (covering 1800 days) obtained with HARPS for HD207129, a G2 star exhibiting a cyclic behavior with a good correlation between the {\it RV} and LogR'$_{\rm HK}$ (0.78).  Spectra and {\it RVs} (hereafter $RV_{\rm drs}$) have both been retrieved from the ESO archives. The spectra were processed as indicated in \cite{meunier17}, and two {\it RV} series were then extracted following the method proposed in this paper, for the sets of lines S$_0$ and S$_1$. The average S/N of the spectra (average over the 72 orders of the echelle spectra from the ESO archive) is around 167. The two series $RV_0$ and $RV_1$ are well correlated, but they show a greater dispersion at small scales than that observed for $RV_{\rm drs}$.  The correlation of $RV_0$ and $RV_1$ with LogR'$_{\rm HK}$ is indeed weaker (around 0.4). The value of $\alpha$ estimated with the different techniques takes very different values, showing that it is not reliable.  Similar conclusion are reached after averaging the data over 50-day bins (the number of spectra per bin is between 1 and 36), confirming that it would not be possible to apply the method on current data unless we  had many more observations. Overall, the noise on $RV_1$ is high, and after correction the time series contain the noise from both $RV_0$ and $RV_1$, which renders the correction impossible for this time series.


\section{Conclusion}

We  tested a new method for correcting for the {\it RV} component due to the inhibition of convection in plages. We use different sets of spectral lines with different depths, whose dependence on the convective blueshift varies. Based on simulated {\it RV} time series, we  identified a set of lines that give performance results in the solar case. We obtained the following results: 

\begin{itemize}
\item{The set of lines must be chosen to provide a convective blueshift as different as possible from the global set of lines while still giving good S/N performance. We found that combining the global set of lines with a set selecting solar lines with fluxes (bottom of the spectral lines) in the range 0.05 -- 0.5 gives good results. The optimal set of lines should be adapted to each star.}
\item{Several methods were tested to reconstruct the parameter $\alpha$ defined as the ratio between the convective blueshift corresponding to the restricted set of lines and the convective blueshift corresponding to the global set of lines. They give similar results overall. One of these methods is quite insensitive to the noise (with the range tested, below 1~m/s), but is biased if the zero of the {\it RV} times series is not precisely known. The other methods are not sensitive to its effect, but are very sensitive to the noise. 
As the different methods are based on different assumptions on the relationship between $RV_{\rm sppl}$ and $RV_{\rm conv}$, it is probably better to test the different techniques for any new {\it RV} time series. 
}
\item{We find a significant improvement at low noise levels, typically below 10~cm/s (for the complete set of lines). For example, for the full temporal sampling (all points except a four-month gap each year), the power in the range 100-500~d  is decreased by 3 orders of magnitude at very low noise levels. Under the conditions considered in this paper it should be possible to reach detection limits at 480~d less than 1 M$_{\rm Earth}$ below 15~cm/s. }
\item{The results remain good with a degraded temporal sampling, although this threshold decreases significantly. The detection limits after correction also increase as the temporal sampling is degraded at all noise levels, but this is not due to the quality of the correction of the convective component, which  also get worse when considering the $RV_{\rm sppl}$ alone. }
\item{We have discussed the impact of two additional types of noise on the {\it RV} time series: the instrumental stability (short timescale) and the granulation (derived from a realistic simulation with an amplitude of 0.8 m/s).  We find that the impact of the instrumental noise is very small for 10 cm/s, and has a small impact at 0.5 m/s. 
The addition of the granulation noise does not impact the performance significantly either, as it behaves as the convective component we focus on in this paper: as the granulation noise is highly stochastic and it is difficult to average out completely, due to the presence of power at large periods and uncorrelated with photometric time series \cite[][]{meunier15}, this method may in principle be a solution to correct for this contribution as well, although other methods have been explored, such as that of  \cite{sulis16}. 
}
\end{itemize}

Our approach allows a correction, based on a physical assumption, of the stellar signal at long timescales (cycle), but also of part of the signal at the rotational period due to the variation of the convective blueshift with activity. 
Other sources of {\it RV} variations of stellar origin remain after this correction, such as the photometric spot+plage component. The performance levels depend strongly on the signal-to-noise ratio: future instruments (such as ESPRESSO/VLT) that allow  very low levels of uncertainties to be achieved on {\it RV} measurements, not only in terms of instrumental stability, but also in term of S/N on the spectra to allow very precise line positions, will therefore be crucial. 




\begin{acknowledgements}
This work has been funded by the ANR GIPSE ANR-14-CE33-0018.
This work has made use of the VALD database, operated at Uppsala University, the Institute of Astronomy RAS in Moscow, and the University of Vienna. This work has made use of the BASS2000 data base at http://bass2000.obspm.fr/.
Our group is part of the LabEx OSUG@2020 (Investissement d'avenir - ANR10 LABX56).
\end{acknowledgements}

\bibliographystyle{aa}
\bibliography{b30328}

\end{document}